\documentclass[12pt,preprint]{aastex}
\usepackage{graphicx,epsfig}
\usepackage{lscape}
\usepackage{rotate}
\usepackage{multirow}
\newcommand{\bq}{\begin{equation}}
\newcommand{\eq}{\end{equation}}

\def\gtsim{\lower.5ex\hbox{$\buildrel > \over\sim$}}
\def\ltsim{\lower.5ex\hbox{$\buildrel < \over\sim$}}

\def\apjl{ApJL}
\def\apj{ApJ}
\def\apjs{ApJS}
\def\mnras{MNRAS}

\def\aap{A\&A}

\def\nat{Nature}

\shorttitle{SN Light Curves}
\shortauthors{Chatzopoulos,Wheeler,Vinko,Horvath,Nagy}
\begin{document}
\title
{ANALYTICAL LIGHT CURVE MODELS OF SUPER-LUMINOUS SUPERNOVAE: $\chi^{2}$-MINIMIZATION OF PARAMETER FITS}
\author{E. Chatzopoulos\altaffilmark{1}, J. Craig Wheeler\altaffilmark{1}, J. Vinko\altaffilmark{1,2}, Z. L. Horvath\altaffilmark{2} 
\& A. Nagy\altaffilmark{2}}
\authoremail{manolis@astro.as.utexas.edu}
\altaffiltext{1}{Department of Astronomy, University of Texas at Austin, Austin, TX, USA.}
\altaffiltext{2}{Department of Optics and Quantum Electronics, University of Szeged, Hungary}

\begin{abstract}

We present fits of generalized semi-analytic supernova (SN) light curve (LC)
models for a variety of power inputs including $^{56}$Ni and $^{56}$Co radioactive decay,
magnetar spin-down, and forward and reverse shock heating due to supernova
ejecta-circumstellar matter (CSM) interaction. We apply our models to the observed
LCs of the H-rich Super Luminous Supernovae (SLSN-II) SN~2006gy,
SN~2006tf, SN~2008am, SN~2008es, CSS100217, the H-poor SLSN-I SN~2005ap,
SCP06F6, SN~2007bi, SN~2010gx and SN~2010kd as well as to the interacting
SN~2008iy and PTF~09uj. Our goal is to determine the dominant mechanism
that powers the LCs of these extraordinary events and the physical conditions
involved in each case. We also present a comparison of our semi-analytical results
with recent results from numerical radiation hydrodynamics calculations in the
particular case of SN~2006gy in order to explore the strengths and weaknesses of
our models. We find that CS shock heating produced by ejecta-CSM interaction
provides a better fit to the LCs of most of the events we examine. We discuss
the possibility that collision of supernova ejecta with hydrogen-deficient CSM
accounts for some of the hydrogen-deficient SLSNe (SLSN-I) and may be a plausible
explanation for the explosion mechanism of SN~2007bi, the pair-instability
supernova (PISN) candidate.
We characterize and discuss issues of parameter degeneracy.

\end{abstract}


\vskip 0.57 in

\section{INTRODUCTION}\label{intro}

The discovery of SLSNe (Quimby et al. 2007; Smith et al. 2007; Gal-Yam 2012; Quimby
et al. 2013) imposed challenges to the widely used mechanism of $^{56}$Ni and $^{56}$Co radioactive
decay diffusion (Arnett 1980, 1982, 1996; hereafter A80, A82, A96) as the typical power
input of many observed SN LCs that do not display prominent plateaus. Attempts to fit the
LCs of some SLSNe provided estimates for the mass of radioactive nickel, $M_{Ni}$, needed to
power the peak luminosity that were close to or far exceeded corresponding estimates for the
total mass of the SN ejecta (Smith et al. 2007; Chatzopoulos et al. 2011, 2012a; see Gal-Yam
2012 for a review). The striking variety in LC shapes, peak luminosities, durations, decline
rates and in spectral evolution makes the determination of a consistent physical model for
the SLSNe even more challenging. Radiation hydrodynamics simulations of interactions of
SN ejecta with massive CSM shells of various power-law density profiles (Moriya et al. 2011;
2012, Ginzburg \& Balberg 2012) provided important insights to the dependence of the main
features of the resulting numerical LCs on the model parameters and were used to reproduced
the observed LCs of some SLSNe (SN~2005ap, SN~2006gy and SN~2010gx).

Chatzopoulos, Wheeler \& Vinko (2012; hereafter CWV12) presented generalized semi-analytical
models for SN LCs that take into account a variety of power inputs such as
thermalized magnetar spin-down and forward and reverse shock heating due to SN ejecta
- CSM interaction with some contribution from $^{56}$Ni and $^{56}$Co radioactive decay. CWV12
considered cases where the photosphere of the diffusion mass is either expanding homologously
or is stationary within an optically-thick CSM. Their formalism was largely based
on that of A80, A82 to incorporate an approximation for radiative diffusion and on that
of Chevalier (1982) and Chevalier \& Fransson (1994; see also Chevalier \& Fransson 2001)
to estimate the luminosity input from forward and reverse shocks depositing kinetic energy
into the CSM and the SN ejecta, respectively. The CSM interaction plus $^{56}$Ni and $^{56}$Co
radioactive decay LC model was succesfully compared in CWV12 to some radiation hydrodynamics
numerical LC models within uncertainties and was used to reproduce the LC of the
SN~2006gy. The availability of easily-computed analytical models allows the development of
a $\chi^{2}$-minimization fitting code that can be used to fit the observed LCs of SLSNe and other
interesting transients. This fitting procedure allows us to estimate the physical parameters
involved and their uncertainties and to assess parameter degeneracies, a severe problem with
multi-parameter models, either analytic or numerical. The resulting models give us a hint
of which power input mechanism is most likely involved in these extraordinary events. This
work serves as a sequel to the work of CWV12 and aims to apply fits of the models presented
there to all SLSNe for which LCs were available when the work was done. The parameters
derived from those fits may be used as a starting point for more accurate, but computationally
expensive numerical simulations in the attempt to understand the physics involved in these rare explosions.

We organize the paper as follows. In \S 2 we summarize the analytical LC models that
were presented by CWV12 and used in this work to fit observed LCs. We also present comparisons
of our semi-analytic SN ejecta-CSM interaction model from CWV12 with numerical
LC models of SN~2006gy. In \S 3 we describe our observational sample of SLSNe and SN IIn
and in \S 4 the fitting method that was incorporated in our $\chi^{2}$-minimization fitting code and
present an analysis of how it calculates uncertainties and parameter degeneracy related to
the large parameter space. We also present model fits to all events in our sample. Finally,
in \S 5 we summarize our conclusions.

\section{SIMPLE MODELS FOR SLSNe LIGHT CURVES}\label{lcmods}

The analytical SN LC models that we use to fit observed SLSN LCs are presented in
detail in CWV12. Here, we give a review of the models and of their physical assumptions.
The derivation of those models was largely based on the methods discussed in A80, A80,
A96 making the assumptions of homologous expansion for the SN ejecta, centrally located
power input source, radiation pressure being dominant and separability of the spatial and
temporal behavior. In the generalized solutions presented in CWV12 we have relaxed the
criterion for homologous expansion of the ejecta and also considered cases for large initial
radius as may be the case for the progenitors of some SLSNe, as well as cases where the
photosphere is stationary within an optically-thick CSM envelope, as may be the case for
luminous interacting SNe IIn. The CSM interaction models include bolometric LCs for both
optically-thick and optically-thin situations, as appropriate. The generalized solutions are
presented for a variety of power input mechanisms including those that have been proposed
in the past.

The first power input mechanism considered is the radioactive decay of $^{56}$Ni and $^{56}$Co
(hereafter the ÒRDÓ model) that leads to the deposition of energetic gamma rays that are
assumed to thermalize in the homologously expanding SN ejecta. As presented in CWV12,
the generalized LC model in this case has the following form:
\begin{eqnarray}
L(t)=\frac{2 M_{Ni}}{t_{d}} e^{-[x^2 + 2 w x]} \{ (\epsilon_{Ni}-\epsilon_{Co}) \times \nonumber \\ 
\times \int_0^x (w + x') e^{[x'^2 + 2 w x']} e^{-t_d/t_{Ni}x'} t_d dx' +\nonumber \\
+ \epsilon_{Co} \int_0^x (w + x') e^{[x'^2 + 2 w x']} \times \nonumber \\
\times e^{-t_d/t_{Co}x'} t_d dx' \} \cdot \left ( 1-e^{-A t^{-2}} \right ),
\end{eqnarray}
where $L(t)$ is the output luminosity in erg~s$^{-1}$, 
$t$ is the time in days relative to the time of explosion $t_{ini}$ (see the Appendix),
$M_{Ni}$ is the initial nickel mass,
$t_{d}$ is the effective light curve time scale the mean of the hydrodinamical and diffusion time scales as defined by A80)
$x = t/t_d$ is the dimensionless time variable
$R_{0}$ is the initial radius of the progenitor,
$v$ is the expansion velocity of the ejecta, 
$w = R_0 / (v t_d)$ is the ratio of the hydrodynamical and the light curve time scales,
$t_{Ni} =$~8.8 days,
$t_{Co} =$~111.3 days are the time scales of Ni- and Co-decay,
$\epsilon_{Ni} = 3.9 \times 10^{10}$~erg~$s^{-1}$~$g^{-1}$
and $\epsilon_{Co} = 6.8 \times 10^{9}$~erg~$s^{-1}$~$g^{-1}$ are the specific energy generation rates
due to Ni and Co decays respectively (Nadyozhin 1994; Valenti et al. 2008). The factor
$(1-e^{-At^{-2}})$ accounts for the gamma-ray leakage, where large $A$
means that practically all gamma rays and positrons are trapped. The gamma-ray optical depth of 
the ejecta is taken to be $\tau_{\gamma} = \kappa_{\gamma} \rho R = At^{-2}$, where $\kappa_{\gamma}$ is
the gamma-ray opacity of the SN ejecta (typically $\sim$~0.03~cm$^{2}$~g$^{-1}$; Colgate, Petschek and Kriese 1980).
Taking into account that $R_{0}/v t_{d} <<$~1 for all SNe considered in this paper, the application of Equation 1
is greatly simplified.
Thus, the fitting parameters for this model are $M_{Ni}$, $t_{d}$ and $A$, 
since reasonable assumptions can be made for $v$ based on observations. 

Another input that has recently been used to explain the LCs of some SLSNe such as SN~2008es and SN~2007bi
is that of the energy released by the spin-down of a young magnetar located in the center of the SN ejecta
(hereafter the ``MAG" model; Ostriker \& Gunn 1971; Arnett \& Fu 1989; Maeda et al. 2007; Kasen \& Bildsten 2010; Woosley 2010). The LC of a SN
powered by such input is given by the following formula:
\begin{equation}
L(t)=\frac{2 E_{p}}{t_{p}} e^{-[x^2 + w x]} 
\int_{0}^{x}  e^{[x'^{2}+ w x]} \frac{x' + w }{(1+yx')^{2}} dx',
\end{equation}
where $x = t/t_{d}$ and $y = t_{d}/t_{p}$ with $t_{d}$ again being an ``effective" diffusion time,
$E_{p}$ the initial magnetar rotational energy and $t_{p} = $ the characteristic time scale for spin-down 
that depends on the strength of the magnetic field.
For a fiducial moment of inertia ($10^{45}$~g~cm$^{2}$), the initial period of the magnetar in units of 10~ms is given by 
$P_{10} = (2 \times 10^{50}$erg/s /$E_{p})^{0.5}$. 
The dipole magnetic field of the magnetar can be estimated from $P_{10}$ and $t_{p}$ as $B_{14} = (1.3 P_{10}^{2}/t_{p,yr})^{0.5}$, 
where $B_{14}$ is the magnetic field in units of $10^{14}$~G and $t_{p,yr}$ is the characteristic time scale for spin-down 
in units of years. Therefore, for the MAG model the fitting parameters are
$E_{p}$, $t_{p}$, $R_{0}$ and $t_{d}$. We note that this model assumes that the input from the pulsar is thermalized
in the ejecta. Simulations of this process show that the energy may not thermalize, but be ejected as magneto-hydrodynamic (MHD) jets
(Bucciantini et al. 2006), thus compromising the mechanism as a model
for SLSNe.

For both the RD and the MAG models, the SN ejecta mass, $M_{ej}$,
is given by the following equation:
\begin{equation}
M_{ej}= \frac{3}{10} \frac{\beta c}{\kappa} v t_{d}^{2},
\end{equation}
where $M_{ej}$ is the mass of the SN ejecta, $\beta$ is an integration constant equal to about 13.8, and $c$ is the speed
of light. The value of $M_{ej}$ for a particular SN determined by its LC is uncertain because of the uncertainties associated
with $\kappa$. For the purposes of this work we will adopt the Thomson electron scattering opacity for fully ionized solar metallicity 
material ($\kappa \sim 0.33$~cm$^{2}$~g$^{-1}$). We also adopt as a
fiducial value for the expansion velocity $v =$~10,000~km~s$^{-1}$ for
the estimates presented in the Tables. The uncertainty of $M_{ej}$ has an important effect on the criterion $M_{Ni} <$~$M_{ej}$
that serves as a consistency check for the RD model.

Another power input that is accepted as being the dominant one in the case of some
SLSNe and SN IIn is that of shock heating. Some SN progenitors are embedded within
dense CSM environments that are formed via continouus or sporadic mass loss. When the
SN explosion occurs, the SN ejecta violently interact with the CSM producing a double shock
structure composed of a forward shock moving in the CSM and a reverse shock moving back
into the SN ejecta. Both the forward and the reverse shocks deposit kinetic energy into the
material which is then radiatively released powering the LCs of these events. The physics of
SN ejecta - CSM interaction is described in the works of Chevalier (1982) and Chevalier \&
Fransson (1994). Simplified models based on this mechanism have recently been considered
as a power source for some SN IIn (Wood-Vasey et al. 2004; Ofek et al. 2010; 
Chevalier \& Irwin 2011). The X-ray flux produced
by CSM interaction-powered events has also been studied in different contexts
(CWV12; Chevaler \& Irwin 2012; Svirski et al. 2012; Ofek
et al. 2013; Pan et al. 2013; Chevalier 2013). In addition, a few numerical radiation
hydrodynamics simulations have been performed yielding broad-band model LCs with the
purpose of reproducing the LCs of some events (Chugai et al. 2004; Moriya
et al. 2011, 2012, 2013). CWV12 presented an analytical LC model that incorporates the
effects of forward and reverse shock deposited energy with those of diffusion through an
optically-thick CSM under the assumption that the shocks are deep within the photosphere
so that the typical shock crossing time scale is larger than the effective radiation diffusion
time scale. The models of CWV12 are in the same regime discussed by Chevalier \& Irwin
(2011) but their generalized model for diffusion spans both cases examined by Chevalier \& Irwin
($R_{d} < R_{CSM}$ and $R_{d} > R_{CSM}$, with $R_{d}$ defined to be the distance from which
radiation can escape from the forward shock). CWV12 also presented a hybrid version
of this model where the effects of energy deposition by the radioactive decays of $^{56}$Ni and $^{56}$Co
are also considered (hereafter the ÒCSM+RDÓ model), given by the following expression:
\begin{eqnarray}
L(t)=\frac{1}{t_{0}} e^{-\frac{t}{t_{0}}} \int_0^t e^{\frac{t'}{t_{0}}} 
[ \frac{2 \pi}{(n-s)^{3}} g^{n\frac{5-s}{n-s}} q^{\frac{n-5}{n-s}} (n-3)^{2} \times \nonumber \\
\times (n-5) \beta_{F}^{5-s} A^{\frac{5-s}{n-s}}
(t'+t_{i})^{\gamma} \cdot \theta(t_{FS,BO}-t') + \nonumber \\
2 \pi (\frac{A g^{n}}{q})^{\frac{5-n}{n-s}} \beta_{R}^{5-n} g^{n} \times \nonumber \\
\times \left ( \frac{3-s}{n-s} \right )^{3} (t'+t_{i})^{\gamma} \cdot \theta(t_{RS,*}-t') ]  dt' + 
\nonumber \\
+ \frac{M_{Ni}}{t_{0}'} e^{-\frac{t}{t_{0}'}} \int_0^t e^{\frac{t'}{t_{0}'}}
\left [ (\epsilon_{Ni}-\epsilon_{Co})e^{-{t' \over t_{Ni}}} + \epsilon_{Co}e^{- {t' \over t_{Co}}} \right ] dt',
\end{eqnarray}
where $t_{0}$ and $t_{0}'$ correspond to the diffusion time-scales through the mass of the optically-thick part of the CSM, $M_{CSM,th}$, and 
the sum of the mass of the ejecta and the optically-thick part of the CSM, $M_{ej}+M_{CSM,th}$, respectively,
$s$ is the power-law exponent for the CSM density profile,
$q=\rho_{CSM,1} r_{1}^{s}$, where $\rho_{CSM,1}$ is the density of the CSM shell at $r=r_{1}$ (we use as a fiducial value
$r_{1} = R_{p}$ where $R_{p}$ the radius of the progenitor star, thus we set the density scale of the CSM, $\rho_{CSM,1}$, immediately
outside the stellar envelope),
$g^{n}$ is a scaling parameter for the ejecta density profile, 
$g^{n} = 1/(4 \pi (\delta-n))[2(5-\delta)(n-5)E_{SN}]^{(n-3)/2}/[(3-\delta)(n-3)M_{ej}]^{(n-5)/2}$, where
$n$ is the power-law exponent of the outer component, and $\delta$ is the slope of the inner density profile of the ejecta 
(values of $\delta =$~0, 2 are typical), $E_{SN}$ is the total SN energy,
$\gamma = (2n+6s-ns-15)/(n-s)$, 
$\beta_{F}$, $\beta_{R}$ and $A$ are constants that depend on the values of $n$ and $s$ and, for a variety of values, are
given in Table 1 of Chevalier (1982), 
$\theta(t_{FS,*}-t)$ and $\theta(t_{RS,*}-t)$ denote the Heaviside step functions that
control the termination of the forward and reverse shock respectively ($t_{FS,*}$ and $t_{RS,*}$ are the termination time scales for
the two shocks) and $t_{i} \simeq R_{p}/v_{SN}$ is the initial time of the
CSM interaction that sets the initial value for the luminosity produced by shocks where $v_{SN} = [10(n-5)E_{SN}/3(n-3)M_{ej}]^{\frac{1}{2}}/x_{0}$ 
is the characteristic velocity of the SN ejecta and $x_{0}=r_{0}(t)/R_{SN}(t)$ is the dimensionless radius
of the break in the SN ejecta density profile from the inner flat component (described by $\delta$) to the outer, steeper component
(described by $n$), which is at radius $r_{0}(t)$. This hybrid CSM+RD model can be easily turned into a pure CSM interaction LC model by 
setting $M_{Ni}=$~0. Moriya et al. (2013) noted that in the case of SN~2006gy CSM interaction is the dominant source of the input energy
with any additional energy due to $^{56}$Ni and $^{56}$Co decays found to be negligible and not to contribute to any of the main features of the resulting LC.

The forward and reverse shock termination time-scales are given by:
\begin{equation}
t_{FS,*}= \left |\frac{3-s}{4 \pi \beta_{F}^{3-s}} q^{{3-n}\over {n-s}} (Ag^{n})^{{s-3} \over {n-s}} \right |^{\xi} 
M_{CSM}^{\xi},
\end{equation}
where $\xi = (n-s)/((n-3)(3-s))$, and
\begin{equation}
t_{RS,*}=\left [ \frac{v_{SN}}{\beta_{R} (Ag^{n}/q)^{\frac{1}{n-s}}} \left ( 1-\frac{(3-n)M_{ej}}{4\pi v_{SN}^{3-n} g^{n}} \right 
)^{\frac{1}{3-n}} \right ]^{\frac{n-s}{s-3}},
\end{equation}
respectively. A simplifying and convenient assumption of our model is that it does not take into account the movement of the shocks in the 
involved diffusion masses (SN ejecta and CSM) which has a direct effect on the LC diffusion time-scale. 
In reality, the forward shock will propagate towards the photosphere within the optically-thick
CSM envelope therefore having an ever-decreasing diffusion time that will lead to a faster evolution for the LC. This caveat was 
discussed in CWV12 and underlined in radiation-hydrodynamic models
recently presented by Moriya et al. (2012). In \S2.1 we present a comparison between our CSM+RD model LC for SN~2006gy and the numerical model obtained by Moriya et al. (2013)
using the same parameters that they used in one of their best-fitting models.

The large parameter space associated with SN ejecta - CSM interaction is reflected by the large number of fitting parameters in this hybrid CSM+RD
model: parameters associated with the nature of the progenitor star ($\delta$, $n$, $v_{SN}$, $R_{p}$, $M_{ej}$, $M_{Ni}$) 
and parameters associated with the nature of the CSM ($M_{CSM}$, $s$,
$\rho_{CSM,1}$). Since we fix $\delta$, $n$ and $s$ for the model fits
presented here, we have a total
of 6 fitting parameters, making fits to observed SN LCs hard to constrain. The main fitting parameters can be used to derive other
physical quantities that give constraints on the configuration of the
CSM envelope implied by a certain fit, in particular the
energy of the supernova explosion $E_{SN} =[3(n-3)/(2(5-\delta)(n-5)]
M_{ej} (x_{0} v_{SN})^{2}$ (where $x_{0}$ the dimensionless radius
where the supernova ejecta density profile breaks from a flat to a
steep power-law)
the radius of the photosphere within the CSM envelope, $R_{ph}$, the total radius of the CSM shell, $R_{CSM}$, and the 
optical depth of the CSM, $\tau_{CSM}$. Due to the large parameter space associated with the hybrid CSM+RD 
LC model, a simplified version of shock heating input, which considers the input to be a ``top-hat" function
of time, is also presented in CWV12 and can be used for some illustrative fits (hereafter the ``TH" model). 
In this model, constant shock energy input $E_{sh}$ is injected
in a diffusion mass for a time, $t_{sh}$, and then shuts off. For this model, the output LC in the case of fixed photospheric radius is given by:
\begin{equation}
L(t) = \cases{\frac{E_{sh}}{t_{sh}} [1-e^{-t/t_{d}}], & $t < t_{sh}$, \cr
\frac{E_{sh}}{t_{sh}} e^{-t/t_{d}} [e^{t_{sh}/t_{d}}-1], & $t > t_{sh}$}.
\end{equation}

\subsection{{\it SN~2006gy: A comparison with results by radiation hydrodynamics calculations.}}\label{moriya}

In CWV12 we presented an indicative fit of the hybrid CSM+RD model to the observed KAIT LC of the archetypical
super-luminous SN~2006gy (Smith et al. 2007), 
and provided a discussion of the event in the context of a variety of LC powering
mechanisms. Moriya et al. (2013) presented 1-D radiation hydrodynamics simulations of SN ejecta with CSM envelopes with power-law
density profiles for several power-law indices ($s =$~0, 2 and 5) performed with the code STELLA (Blinnikov \& Bartunov 1993). 
Moriya et al. (2013) also present a comparison of their numerical LC with our analytical model
using the same parameters as presented in CWV12 and note several discrepancies between the two (their section 6.4 and Figure 14). We therefore find it interesting
to further compare our approximate models with their numerical results for SN~2006gy in order to better assess the limitations and weaknesses due to the
assumptions made for the analytical models. 

To do so, we pick the parameters given for model F1 of Moriya et al. (2013), which is one of their models that best reproduces the observed LC of SN~2006gy.
For this model we have: $E_{SN} =$~$10^{52}$~erg, $M_{ej} =$~20~$M_{\odot}$, $M_{CSM} =$~15~$M_{\odot}$, 
$\delta =$~1, $n =$~7, $s =$~0 and $M_{Ni} =$~0.
The outer radius of the CSM is $R_{CSM,o} =$~$1.1 \times 10^{16}$~cm and the inner radius $R_{CSM,i} =$~$5 \times 10^{15}$~cm
giving a thickness of the CSM shell of $\Delta R =$~$6 \times 10^{15}$~cm. 
While our radii are different, our $\Delta R$ is the same as Moriya et al. (2013).
Thomson scattering is the dominant source of opacity in the CSM beyond the forward shock
in the simulations of Moriya et al. (2013) and has the value $\kappa =$~0.34~cm$^{2}$~s$^{-1}$ for a solar mixture ($X =$~0.7), but it is calculated
self-consistently in the radiation hydrodynamics calculations. The radius of the progenitor
star is not a parameter considered in the simulations of Moriya et al. (2013) who start their simulations by considering freely expanding SN ejecta with a density
profile that is described by Chevalier \& Soker (1989). It is likely that the progenitors of those SNe are large, therefore we adopt $R_{p}$ to be in
the range $\sim$~$10^{13}$-$10^{14}$~cm consistent with either blue supergiant stars (BSG) or RSG stars. Moriya et al. (2013)
consider the collision between the SN ejecta and the CSM shell to be inelastic, therefore associated with an energy conversion efficiency. For the F1 model, it is
determined that only 29\% of the total SN energy is converted to radiation yielding $E_{rad} =$~$2.9 \times 10^{51}$~erg. In our semi-analytical
hybrid CSM+RD model the conversion efficiency of the shock kinetic energy to radiation is assumed to be 100\%.
Additionally, in order to take into account multi-dimensional effects such as Raleigh-Taylor instabilities (present in the dense CSM shell) 
in 1-D calculations Moriya et al. (2013) consider a ``smearing" parameter $B_{q}$. 
Taking these model variations into consideration together with the uncertainty associated when converting 
the observed magnitudes to bolometric luminosities induce a general uncertainty
in the value of $E_{SN}$. Moriya et al. (2013) estimate $E_{SN}$ to be greater than $4 \times 10^{51}$~erg for SN~2006gy. 

Keeping in mind the above-mentioned uncertainties in the values of $E_{SN}$, $R_{p}$ and $\kappa$ we present several variations of the model F1 LC in Figure 1
and Table 1
as calculated with our semi-analytical CSM interaction model presented by Equation 4
in order to explore how closely our models agree with theirs for similar parameter choices.
In the top left panel of Figure 1 we present model C1 with 
$E_{SN} =$~$2.2 \times 10^{51}$~erg, $R_{p} =$~$10^{13}$~cm, $R_{CSM} =$~$2.4 \times 10^{15}$~cm and 
$\kappa =$~0.2~cm$^{2}$~s$^{-1}$. This value of $\kappa$ is suitable for a hydrogen-poor CSM. Although this is not the case for SN~2006gy, this
choice allows for the fact that one of the assumptions of the semi-analytical model is that the energy deposition from the forward and reverse shocks 
takes place at a constant radius deep within the CSM while, in reality, the double-shock structure moves outwards in radius reaching smaller and smaller 
optical depths and resulting in ever-decreasing diffusion time-scales. As a result, a way to account for this diffusion time-scale decrease in our
model is to assume a smaller ``effective" optical opacity. 

It can be seen in Figure 1 that this model represents well the rising part and the peak luminosity 
of the LC of SN~2006gy, but has difficulty fitting the post-maximum decline. A much better result is obtained for an even smaller 
choice for $\kappa$ (0.09~cm$^{2}$~s$^{-1}$) and for 
$E_{SN} =$~$1.7 \times 10^{51}$~erg, $R_{p} =$~$0.9 \times 10^{13}$~cm and $R_{CSM} =$~$2 \times 10^{15}$~cm (upper right panel, model C2).
In the lower left panel we present three more variations of the Moriya et al. (2013) F1 model that use the same $E_{SN} =$~$10^{52}$~erg, $R_{p} =$~$10^{14}$~cm
and $\kappa =$~0.33~cm$^{2}$~s$^{-1}$, but with varying slope of the ejecta density profile and the mass of the CSM
($n =$7, $M_{CSM} =$~15~$M_{\odot}$ for the red curve fitted in the open circles (model C3); $n =$12, $M_{CSM} =$~15~$M_{\odot}$ for the green curve fitted in the 
open squares (model C4);
$n =$12, $M_{CSM} =$~5~$M_{\odot}$ for the blue curve fitted in the open triangles (model C5)). The open circles, squares and triangles represent the same SN~2006gy LC
data moved in the time axis by different constant values (days) in order to best match the corresponding models.
We note that for these models we had to scale down the resulting luminosities
by factors of 5-7 in order to fit the SN~2006gy data. As we noted above, this uncertainty results from the fact that we assume 100\% conversion
efficiency from kinetic energy of the shocks to radiation in our model (leading to more luminous outputs). 
Decreasing the luminosity is roughly equivalent to decreasing the conversion efficiency. 
Also, we recall that our LC of SN~2006gy 
is constructed assuming a zero BC for the observed magnitudes. The BC for such a complex luminous SN IIn is expected to be large and to vary
with time as the LC evolves, especially due to the fact that the bulk of the shock deposited energy is emitted at short wavelengths (UV and soft X-rays)
particularly in early epochs (Chevalier \& Fransson 1994). It can be
seen that model C4 best reproduces the LC of SN~2006gy. This model uses the
same parameters as the F1 model of Moriya et al. (2013) with a different choice for $n$ (12 instead of 7 that corresponds better 
to the SN ejecta density profile slope for an RSG progenitor star) and the luminosity scale-down by a factor of $\sim$~7.

The comparisons of the semi-analytical versions of the F1 model of Moriya et al. (2013) with the observed LC of SN~2006gy presented above lead to the
main conclusion that, within the uncertainties associated with the semi-analytic model and its simplifying and convenient assumptions, it can be a useful
tool to provide estimates of the physical parameters associated with SLSN-II. The considerable differences between the first version of
our hybrid CSM+RD model for SN~2006gy that we presented in CWV12 and the numerical LC of Moriya et al. (2013) for the same parameters are partially
attributable to the large parameter space associated with SN ejecta - CSM interaction that makes it hard to find an ``absolute" minimum value for $\chi^{2}$
representing the true best-fit to the LC data. It is possible that there are several combinations of the semi-analytic CSM interaction model 
parameters that produce fits of similar quality but for which more accurate, 
numerical LCs might not be good representations of the observed LC. This was the case for the initial CSM+RD model we presented for the LC of SN~2006gy
in CWV12.
For this reason we think it is useful to use the semi-analytical models in order to obtain a number of good fits corresponding to $\chi^{2}$ minima and then
to use those fits as starting points to perform more computationally expensive and physically self-consistent numerical radiation hydrodynamics simulations 
that will certainly clarify which parameter choice best matches observed LC data.

\section{THE OBSERVATIONAL SAMPLE OF SLSNe}

In this section we give a brief description of the SNe studied. We use the available photometric
observations of recently discovered SLSNe that are of spectral type IIn (SN~2006gy,
2006tf, 2008am, 2008es, and CSS100217), as well as those in the hydrogen-deficient category
defined by Quimby et al. (2011) (SN~2005ap, SCP06F6, 2010gx) and those that are candidates
to be PISNe and similar events (SN~2007bi, 2010kd). A recent review by Gal-Yam
(2012) classifies SLSNe in a similar manner, referring to hydrogen-rich events as SLSN-II
and to hydrogen-poor events as SLSN-I, but also defines the SLSN-R category for events
that are thought to explode due to the pair-instability mechanism and hypothesized to be
powered by large amounts of radioactive $^{56}$Ni (SN~2007bi). In our classification scheme, the
SLSN-I category includes the SLSN-R events. We also fit two recent Type IIn SNe (2008iy
and PTF~09uj) that are not SLSNe, but their observed spectra and LCs are governed by
strong CSM interaction. The reason for their inclusion in the present paper is that they can
serve as test cases for our simplified CSM-interaction model described in \S2. A summary
of the basic characteristics of the sample of SNe studied in this work is presented
in Table 2. The black-body temperatures, $T_{BB}$, are estimated using the 
observed peak pseudo-bolometric luminosities and assuming homologous
expansion up to the time of maximum light, therefore using the radius $R_{SN,max} = v_{SN} t_{rise}$
where $v_{SN}$ is the estimated photospheric velocity of each event as derived from spectroscopic
observations. Note that the $T_{BB}$ values
are similar between SLSN-I and SLSN-II (10,000-20,000 K).

\subsection{Normal SNe with strong CSM-interaction}

\subsubsection{SN~2008iy}\label{08iy}

The Catalina Real-Time Transient Survey (CRTS; Drake et al. 2009a) discovered SN IIn 2008iy. SN~2008iy was not a SLSN, but 
had the longest rise time to maximum luminosity known in the history of SNe ($\sim$~400d; Miller et al. 2010). 

Miller et al. (2010) present an extensive photometric study of SN~2008iy in the IR (PAIRITEL), optical (Nickel and DS) and UV ({\it Swift}) bands. 
Studying
the pre-explosion CRTS frames, Miller et al. (2010) accept the explosion date to be $MJD_{expl} =$~54356, and allow for an uncertainty of
approximately 50 days prior to that. Keck LRIS and Kast Lick-3m Shane telescope spectra
confirmed SN~2008iy as a classic SN IIn with strong intermediate-width H and He emission features. The characteristic
velocity implied by the FWHM of the H$\alpha$ line is $\sim$~5,000~km~s$^{-1}$. 
Miller et al. (2010) marginally detect P Cygni profiles associated with late-time ($\sim$~911~d after discovery) 
H$\alpha$ features that give a hint of photospheric expansion associated with SN~2008iy. The redshift of SN~2008iy is $z =$~0.0411. 
The extremely long rise time of SN~2008iy prompted Miller et al. (2010) to adopt a scenario of extensive CSM interaction as a natural explanation
for this event. They specifically discussed a model of interaction with CSM clumps (Chugai \& Danziger 1994) in which the number density of the clumps increases
over a radius of $\sim 1.7 \times 10^{16}$~cm from the progenitor.

To produce the pseudo-bolometric LC of the event we convert the available DS and Nickel I-band magnitudes of SN~2008iy
to bolometric luminosities, assuming BC=0. Note that the DS band is similar to the SDSS i$^{\prime}$ band, so also in good agreement with Nickel I-band. Using only single-band magnitudes to estimate the bolometric LC is a very approximate
approach, but our intention is to get only order-of-magnitude estimates of the basic physical parameters that affect
the LC by most, without attempting a detailed fine-tuned analysis and modeling of a particular object.

\subsubsection{{\it PTF~09uj}}\label{09uj}

The Palomar Transient Factory reported the discovery of the transient PTF~09uj that was identified as a Type IIn event (Ofek et al. 2010).
This object is not a SLSN but has been modeled with CSM interaction, so we include it in our own study. 
PTF~09uj was discovered during its rise to maximum light by the Oschin 48-inch Schmidt telescope (P48) at Palomar Observatory. Pre-explosion
images by GALEX constrained the explosion time of the SN to be $MJD_{expl} =$~55000, which was used for the LCs presented by Ofek et al. (2010).
P48 R-band and P60 r-band follow up photometry of the event was presented in the same work. A Lick spectrum obtained around peak
luminosity revealed emission lines of H and He, typical for Type IIn events. Only a hint of P Cygni absorption associated with H$\alpha$ was
detected (Ofek et al. 2010). The Lick spectrum was also used to determine the redshift of the SN
($z =$~0.065). Ofek et al. (2010) interpreted PTF~09uj with a model of ejecta-CSM interaction, where the CSM is a dense wind ($s= $~2). The
same study derived the characteristic values $\dot{M} =$~0.1~$M_{\odot}$~yr$^{-1}$ and $v_{w} =$~100~km~s$^{-1}$ for the mass-loss rate
and the velocity of the wind of the progenitor star. Under these assumptions the whole LC of this event is powered by CS shock breakout 
from the optically-thick part of the wind. To derive these estimates, Ofek et al. (2010) took the diffusion time to be equal to the time
of shock-break out and assumed a value $v_{sh} =$~10,000~km~s$^{-1}$ as the typical velocity of the CS shock, which we will also adopt for 
the purposes of our study. As above, we converted the P48 R and P60 r-band LC of PTF~09uj to produce a pseudo-bolometric LC assuming BC=0.

\subsection{{\it Hydrogen-rich super-luminous events (SLSN-II)}}\label{SNIIn}

\subsubsection{SN~2006gy}\label{06gy}

The archetypical SLSN IIn SN~2006gy was discovered by the Texas Supernova Search
(TSS) project and first presented by Smith et al. (2007). At the observed redshift of
SN~2006gy, z = 0.074, the absolute visual peak magnitude of the event reached $\sim$ -22 mag,
making it one of the brightest explosions ever discovered. A rich database of optical spectra
were obtained for SN~2006gy (Smith et al. 2007, 2008, 2010) that provides an extensive
record of its spectral evolution. SN~2006gy showed strong Balmer emission features with
their narrow components associated with P Cygni absorption indicative of photospheric
(or CSM) expansion. The H$\alpha$ line profile evolved throughout the course of the LC of
SN~2006gy showing an evolution that is marked by three phases described in Smith
et al. (2010). The full width at half-maximum (FWHM) of H, around maximum light
reveals characteristic velocities of $\sim$ 4,000~km~s$^{-1}$. Here, we consider the KAIT LC of
SN~2006gy presented in Smith et al. (2007) converted to a pseudo-bolometric LC assuming
bolometric correction BC=0. We also adopt $E(B ? V) =$~0.72 mag yielding R-band
extinction $A_{R} =$~1.68~mag, also in accordance with Smith et al. (2007). It has been suggested
(Smith et al. 2007; Smith \& McCray 2007, Smith et al. 2010) that the progenitor
of SN~2006gy was most likely an LBV-type star that suffered extreme mass-loss prior to its
death. In the same framework, upon explosion the SN ejecta violent collided with the LBV
nebula producing the observed high luminosity via shock energy deposition. A similar model
of interaction between multiply ejected shells in the context of a pulsational pair instability
supernova (PPISN) has also been suggested (Heger \& Woosley 2002; Woosley, Blinnikov \&
Heger 2007).

\subsubsection{SN~2006tf}\label{06tf}

SN~2006tf was a strongly interacting Type IIn SN discovered by the TSS project (Smith et al. 2008). The B,V,R and I band LC of 
SN~2006tf was constructed using observations from the KAIT telescope. The SN was discovered after peak 
luminosity so the explosion date of SN~2006tf remains uknown. This will have an impact on the parameters of the models that we attempt to fit below.
As an initial value for the explosion date, we adopt $MJD_{expl} =$~54050 which is 50 days prior to the first photometric observation by the KAIT
R-band and within the range proposed by Smith et al. (2008). 
The spectra of SN~2006tf were characteristic of the Type IIn subclass showing strong intermediate-width emission features of H. The H$\alpha$
features had FWHM$\simeq$~2,000~km~s$^{-1}$. We stress that is value does not directly reflect a characteristic fluid velocity value
due to the the ambiguity of the interpretation of the line widths which are affected by several broadening mechanisms (see Smith et al. 2012; CWV12).
SN~2006tf exhibited spectroscopic similarities 
with other SLSNe such as SN~2006gy (Smith et al. 2007) and SN~2008am (Chatzopoulos et al. 2011) and, given the duration of its observed LC,
it is considered a classic example of a Type IIn event associated with a massive progenitor (Smith et al. 2008).
In this study we use the available KAIT R-band LC of SN~2006tf converted to bolometric luminosity (BC=0).

\subsubsection{{\it SN~2008am}}\label{08am}

SN~2008am was another bright explosion discovered by the ROTSE Supernova Verification Project (RSVP) (Chatzopoulos et al. 2011). 
The ROTSE-IIIb telescope followed up SN~2008am photometrically for over $\sim$~150d. 
The spectra determine the redshift of
the event to be $z =$~0.2338. For standard $\Lambda$-CDM cosmology this redshift translates to a distance of $\simeq$~1121~Mpc making SN~2008am
one of the most luminous explosions discovered with peak ROTSE magnitude -22.3~mag. The spectra show classic Type IIn Balmer and HeI emission lines
with typical FWHM $\sim$~1,000~km~s$^{-1}$. 
As for the case of SN~2006gy and SN~2006tf, the width of the emission lines may be attributable to electron scattering effects and 
not to true bulk kinematic motion. Chatzopoulos et al. (2011) used the ROTSE LC of SN~2008am to determine the explosion date 
of $MJD_{expl} =$~54438.8~$\pm$~1.

For the purposes of this work, we analyze the ROTSE LC
which has many data points and contains data on the rise. This use of the ROTSE LC is in accord with
the analysis we did for other events discussed here. We convert the ROTSE LC to a pseudo-bolometric one assuming BC=0. 

\subsubsection{SN~2008es}\label{08es}

Another bright SLSN discovered by ROTSE-IIIb and the RSVP program was SN~2008es (Gezari et al. 2009). The ROTSE-IIIb
telescope managed to capture this event before maximum light which allowed Gezari et al. (2009) to constrain the explosion
date of the event ($MJD_{expl} =$~54574~$\pm$~1). 
Post-maximum photometric observations of SN~2008es were obtained in IR (PAIRITEL; Miller
et al. 2008), optical (KAIT, Nickel, PFC, UVOT; Miller et al. 2008 and P60, P200; Gezari et al. 2009) and UV (UVOT; Miller et al. 2008)
bands which allowed a comprehensive study of the event. SN~2008es was followed up spectroscopically for over $\sim$~100~d and exhibited a slow
spectroscopic evolution with nearly featureless spectra in the first $\sim$~20~d after maximum with H$\alpha$ emission appearing only in the nebular
spectra. This, together with the approximately linear decline of the optical LC (in magnitude scale), led Miller et al. (2008) to classify 
SN~2008es as a Type IIL explosion. The spectra revealed the redshift of the SN to be $z =$~0.213 (Miller et al. 2008) with characteristic
velocities of $\sim$~10,000~km~s$^{-1}$ which is the value we use for reference here. P Cygni features were detected in the nebular spectra
for the H and He emission lines and became more prominent as the event evolved, indicating photospheric expansion.
Although SN~2008es did not show classic SN IIn features, CSM interaction was considered as the most likely candidate 
for the event by Miller et al. (2008) and Gezari et al. (2009). They
argue that an initially dense CSM can account for the absence of characteristic CSM interaction emission features. Kasen \& Bildsten (2010)
considered a magnetar model as an explanation for SN~2008es which, although it provides a good fit to the LC, may have difficulty 
in accounting for the spectroscopic features (but see Dessart et al., 2012 for a different conclusion).
We use the P60 r-band together with the ROTSE unfiltered observations to assemble the pseudo-bolometric LC of of SN~2008es.

\subsubsection{CSS100217:102913+404220}\label{CSS}

CSS100217:102913+404220 (hereafter CSS100217) was discovered on February 17, 2010 (Drake et al. 2011). 
Drake et al. (2011)
determine the redshift of the host of CSS100217 spectroscopically to be 0.147, implying a distance of 680.4~Mpc for the event assuming 
standard $\Lambda$-CDM cosmology. At this distance, and assuming Milky-Way extinction of $E(B-V) =$~0.1426 at the position of the SN, the absolute magnitude
of CSS100217 is $M_{V} =$~-22.7 approximately 45d after the discovery corresponding to an optical luminosity of $1.3 \times 10^{45}$~erg~s$^{-1}$
making the event one of the most luminous ever discovered. Multi-wavelength photometry was obtained though the course of the transient in the radio, near-IR,
optical and UV. The transient was also detected by the {\it Swift} XRT in the 0.2-10~keV band as a soft source.
The photometric data yield a total radiated energy of 
$\sim$~$1.2 \times 10^{52}$~erg over a period of $\sim$~287 rest-frame days. The spectra of CSS100217 are similar
to those of SN~2008iy and other Type IIn SN and show narrow Balmer emission lines indicating that the mechanism that powered that event is most likely strong
interaction of SN ejecta with a massive CSM medium. Drake et al. (2011) also consider alternative scenaria such as AGN variability or a tidal disruption event
(TDE) which they rule out based on arguments related to the spectroscopic evolution of the event. 

We convert the CSS LC of CSS100217 (which corresponds to V-magnitudes) to bolometric in accord with Equation 1 of Drake et al. (2011) 
and the assumptions discussed therein. The conversion yields a peak luminosity of $\sim$~$4 \times 10^{44}$~erg~s$^{-1}$, consistent with the one cited
by Drake et al. (2011), at $\sim$~73~d after explosion in the rest-frame assuming $MJD_{expl} =$~55160, the time that the first real detection of the transient
was recorded.

\subsection{{\it Hydrogen-deficient super-luminous events (SLSN-I)}}\label{HdefSN}

\subsubsection{SN~2005ap}\label{05ap}

SN~2005ap was the first SLSN discovered by the Robotic Optical Transient Search Experiment (ROTSE) telescopes
of the TSS program (Quimby et al. 2007). The only available LC of SN~2005ap was that taken
from the ROTSE-IIIb telecope. Even though the S/N ratio was moderate, the post-maximum evolution of the LC of SN~2005ap
shows a fast decline. The exact explosion date of the event is not well-constrained, but Quimby et al. (2007) adopt
a value 7-21 days before maximum based on comparisons with SN IIL template LCs. 
Spectra of SN~2005ap showed broad P Cygni features of C, N and O
that correspond to a velocity of $\sim$~20,000~km~s$^{-1}$. The spectra also indicate a redshift of $z =$~0.2832 for
SN~2005ap, which means that the peak absolute unfiltered magnitude of SN~2005ap was -22.6~mag. Quimby et al. (2011) put SN~2005ap
in the same category as the recently discovered peculiar transients SCP06F6 (Barbary et al. 2008), 
PTF09cwl, PTF09cnd and PTF09atu. 
Recently, Ginzburg \& Balberg (2012) presented a SN ejecta-CSM interaction scenario for SN~2005ap in which the LC of the event
is the result of the violent collision between equal mass ($\sim$~15~$M_{\odot}$) SN ejecta and steady-state wind ($r^{-2}$) CSM. 
Here we consider an $s =$~0 constant density CSM shell instead that may be more consistent with episodic, LBV-type mass-loss that
is implied by the high mass-loss rate suggested for the event. 
Again, we convert the SN~2005ap ROTSE LC to rest-frame pseudo-bolometric LC by assuming zero bolometric correction (BC=0).

\subsubsection{SCP06F6}\label{06f6}

SCP06F6 was a controversial transient discovered by Barbary et al. (2009). The LC of SCP06F6 was constructed by observations
in the F850LP (similar to SDSS z$^{\prime}$) and F775W (similar to SDSS i$^{\prime}$) filters of the Advanced Camera for Surveys (ACS) 
Wide-Field Camera mounted on the Hubble Space Telescope (HST). The LC of SCP06F6 is fairly symmetric in shape with a 
rise time-scale close to its decline time-scale. The redshift of SCP06F6 remained unknown for over two years due to its peculiar
spectrum. Three optical spectra obtained with the Very Large Telescope (VLT) Low Dispersion Spectrograph 2 (FORS2), 
Keck-LRIS and Subaru-FOCAS showed
unidentified broad absorption features in the blue. Works by Gaensicke et al. (2009), Soker et al. (2010) and Chatzopoulos
et al. (2009) made attempts to identify the nature of these features and determine the redshift of SCP06F6 unsuccesfully.
Quimby et al. (2011) associated the spectrum of SCP06F6 with spectra of other similar recently discovered Palomar Transient Factory
(PTF) transients (PTF09cwl, PTF09cnd and PTF09atu) and determined the redshift of SCP06F6 to be $z =$~1.189. The same work
identified the controversal broad spectral features as Fe/Co blends and \ion{Ni}{3}, \ion{O}{2} and \ion{Si}{3} lines.

Using the observed broad-band LCs of SCP06F6, we construct the rest-frame pseudo-bolometric LC following the technique described 
in Chatzopoulos, Wheeler \& Vinko (2009) for the currently accepted redshift of the event ($z = 1.189$, Quimby et al. 2011). 

\subsubsection{SN~2007bi}\label{07bi}

Gal-Yam et al. (2009) reported the discovery of the best candidate for a PISN explosion to date, SN~2007bi, by the Supernova Factory
(SNF) program (Aldering et al. 2009). Palomar-60 (P60) R-band photometry of SN~2007bi was obtained for over $\sim$~130~d period at the rest
frame of the SN. The explosion date of SN~2007bi is uncertain which induces an uncertainty in the models applied to explain the LC of the event.
We adopt as a reference value $MJD_{expl} =$~54089, which is 70 days before peak R-magnitude, in accordance with the range proposed by 
Gal-Yam et al. (2010). The explosion date will be a fitting parameter for the LC of this SN in our work. The spectra of SN~2007bi do not
show signs of CSM interaction and H and He features are not detected. Strong Ca, Mg and Fe features and Ni/Co/Fe blends
are identified close to the NUV part of the spectrum (Gal-Yam et al. 2010). Spectral fits provide us with an estimate of the characteristic
velocity of $v =$~12,000~km~s$^{-1}$.
The lack of H and He features led to the classification
of this event as a SN Ic explosion. Also, the lack of evidence for a SN IIn CSM interaction and the long duration and 
high luminosity of the event make it a good candidate for a PISN explosion (Gal-Yam et al. 2010). Other proposed models for SN~2007bi
are an energetic core-collapse explosion (Moriya et al. 2010) and a magnetar spin-down model developed by Kasen \& Bildsten (2010)
and Woosley (2010). 
Recently, the possibility of H-poor CSM interaction as
a model for SLSN-I events, including SN~2007bi has also been suggested (Chatzopoulos \&
Wheeler 2012b). We constructed the pseudo-bolometric LC of SN~2007bi using the available
P60 R-band LC of SN~2007bi presented in Table 3 of Gal-Yam et al. (2009), a Milky-Way
extinction of $A_{R} =$~0.07 mag from the NED database at the location of the transient, and
the observed redshift $z =$~0.1279. All these values yield an absolute peak luminosity of
$1.11 \times 10^{44}$~erg~s$^{-1}$ for SN~2007bi.

\subsubsection{SN~2010gx}\label{10gx}

SN~2010gx was discovered on March 13, 2010 at 18.5~mag by the CRTS team (Mahabal et al. 2010; Pastorello et al. 2010). Independent discovery
wes later announced by the PTF survey (Quimby et al. 2010). The host of SN~2010gx is identified as a faint galaxy in the SDSS images and its
redshift is estimated to be 0.23. For standard $\Lambda$-CDM cosmology this redshift translates to a distance of $\simeq$~1120~Mpc making this object
yet another member of the class of SLSNe ($M_{B,peak}\simeq$~-21.2). 
Extensive photometric follow up was obtained in the {\it u}{\it g}{\it r}{\it i}{\it z} bands using a variety
of telescopes (Pastorello et al. 2010) which provided an estimate for the explosion date of the event to be $MJD_{expl} =$~55260 yielding a rise time
to maximum light of $\sim$16~d in the rest-frame.

The pre-maximum spectra of SN~2010gx show a blue continuum ($T_{BB} =$15,000$\pm$1700~K) with broad absorption features in the bluer parts. 
Later spectra also show broad P Cygni absorptions of \ion{Ca}{2}, \ion{Fe}{2} and \ion{Si}{2} which led to classification of 
the object as a SN Ib/c. Pastorello et al. (2010)
have difficulty suggesting a model that accounts for the overall characteristics of SN~2010gx (spectral evolution, fast evolution of LC, peak luminosity)
and suggest that most scenarios ($^{56}$Ni decay powered core-collapse SN, PISN, PPISN or magnetar-powered SN) do not comfortably match the event. Therefore
here we attempt to re-visit those models in more detail and also consider CSM interaction as an alternative. SN ejecta interaction with a dense $r^{-2}$ wind
as the power mechanism for SN~2010gx was recently considered by Ginzburg \& Balberg (2012) 
who presented hydrodynamics simulations of such phenomena that take into account the effects of radiation diffusion and calculate model LCs. 
Ginzburg \& Balberg determined that collision of $\sim$~15~$M_{\odot}$ of SN ejecta (with energy $E_{SN} =$~$2 \times 10^{51}$~erg) 
with $\sim$~15~$M_{\odot}$ of a steady-state CSM wind that terminates at a radius of $2.5 \times 10^{15}$~cm reproduced well the observed LC and
black-body temperature evolution of SN~2010gx. The implied average mass-loss rate for their parameters assuming a 
fiducial wind velocity of 10~km~s$^{-1}$ is $\dot{M} =$~0.2~$M_{\odot}$~yr$^{-1}$ which is inconsistent with steady-state, quiescent 
mass-loss and more in accord with episodic, LBV-type mass loss that, in turn, does not necessarily lead to a $r^{-2}$ density profile for the CSM. Here
we will consider CSM interaction with an $s =$~0 constant density CSM shell that might be more consistent with non-steady mass-loss. 

We converted the {\it r} band LC of SN~2010gx to pseudo-bolometric assuming BC=0, $E(B-V) =$~0.04 (Schlegel et al. 1998) and 
$E(B-V)_{host} =$~0 (Pastorello et al. 2010).

\subsubsection{{\it SN~2010kd}}\label{10kd}

The ROTSE-IIIb telescope of the RSVP project discovered SN~2010kd on November 14, 2010 (Vinko et al. 2010) at a magnitude of $\sim$~17~mag. Spectra
obtained by the HET and Keck showed narrow H$\alpha$ emission which helped constrain the redshift of the object at $z =$~0.101 implying an absolute ROTSE
magnitude of $\sim -$21 suggesting the event is super-luminous. The transient was followed up photometrically in the UBVRI filters and in the UV and X-ray
by {\it Swift} UVOT and the XRT. The SN was detected as a strong UV source, but no X-ray flux was measured. 
The photometric observations and the date of discovery provide an estimate for the actual
explosion date at $MJD_{expl} =$~55483 implying a rest-frame rise time to peak luminosity of $\sim$~60d (Vinko et al. 2010; Vinko et al. 2013 in preparation).
The observed spectroscopic evolution of SN~2010kd implies a  behavior similar to SN~2007bi
with a lack of H and He features and presence of \ion{C}{2}, \ion{O}{1}, \ion{O}{2} and possibly \ion{Co}{3}
making it a SLSN-I event, and a PISN candidate. 
We converted the V-band LC of SN~2010kd to pseudo-bolometric assuming BC=0, $E(B-V) =$~0.0213 (Schlegel et al. 1998) which implies a peak luminosity of $\sim$~10$^{44}$~erg~s$^{-1}$.

\section{FITS TO OBSERVED LIGHT CURVES OF SLSNe}\label{obscomp}

In this section we attempt to fit the semi-analytical LC models that are 
given by Equations 1, 2, 4 and 7 to the observed LCs of some interesting SN events, in order to understand which mechanism best 
describes their nature. 

\subsection{{\it The fitting method}}\label{fitting}

The semi-analytic LC models described in \S 2 were fitted to the observed data by applying
a $\chi^2$-minimization code {\tt MINIM} that was developed by two of us (ZLH and JV) and 
implemented in {\tt C++}. {\tt MINIM} uses a controlled
random-search technique, the Price algorithm (Brachetti et al. 1997), which has been extensively tested and applied
for solving global optimization problems.  
The algorithm treats the unknown parameters as random variables
in the $N_p$-dimensional hyperspace, where $N_p$ is the number of parameters. The boundaries of the
permitted values for each parameter are defined as $p_i (low)$ and $p_i (high)$ and given to the code
in the input file. The aim of the algorithm is to find the global minimum of 
the $\chi^2$ function within this permitted parameter volume.  

After reading the input data, the code randomly selects $N_r$ vectors in the parameter hyperspace. Each
vector is defined as $\vec{p}$ = ($p_1$, $p_2$, ..., $p_{N_p}$), and each $p_i$ parameter is chosen
as an equally-distributed random number between $p_i (low)$ and $p_i (high)$. For each $\vec{p}$ vector
the value of $\chi^2$ is calculated.  We have applied
$N_r = 200$, which was found to be a good compromise between reliable convergence (i.e. finding
the global minimum $\chi^2$) and computation speed. The algorithm then chooses a new trial
$\vec{p}$ vector by combining a randomly chosen subset of $N_p + 1$ elements from the stored vectors, 
and compares their $\chi^2$ value with those of the stored vectors. If a vector with a better $\chi^2$ is 
found, then the one with the highest $\chi^2$ in the stored vector set is replaced by this new vector. 
This process is iterated until the difference between the $\chi^2$ values of the stored vectors are 
less than $\Delta \chi^2 = 1$.

As a final step, the Levenberg-Marquardt algorithm is applied to fine-tune the parameters in 
the lowest $\chi^2$ vector produced by the Price algorithm. The result of this routine is accepted
as the best-fitting model parameter set. The uncertainties of the fitted parameters are estimated
by calculating the standard deviation of each $p_i$ parameter in the final set of the $N_r$ random 
vectors around the best-fitting vector. Our tests with simple analytic functions and 
simulated data have shown that this kind of error estimate is consistent with using 
the full covariance matrix of the $\chi^2$ hypersurface. An extensive discussion of parameter correlation and degeneracy
as calculated by {\tt MINIM} in the case of the CSM+RD model is presented in the Appendix.

We note that in some cases the fitting is ill-constrained due to the small number of data points and the presence of many fitting parameters, resulting in a low value of the degree-of-freedom $N_{data}-N_p$.
Estimates for the diffusion or SN ejecta mass, $M_{ej}$, provided by Equation 3 for the RD, MAG and TH models, are uncertain because of the uncertainty in the optical opacity, $\kappa$, 
but also due to the intrinsic uncertainty in the fitted value for the LC time-scale, $t_{d}$, which is dependent
upon the explosion time, $t_{ini}$ (which is also ill-constrained for SLSNe).

For the hybrid CSM+RD model
we assume and fix $\delta =$~2 and $n =$~12 for the SN ejecta inner and outer power-law density profile slopes, respectively,
in all cases. 
The range in fitted parameters for various models that give approximately equally good fits
yield some notion of the range of parameters in viable model space. Ideally, we would incorporate the density slope of the CSM, $s$, as a parameter
to be fitted. This can be done, but substantially increases the computation time. Here we have investigated two relevant values of $s$, 0 and 2, and randomly
varied other parameters.

A summary of the fitting parameters is given in Tables 3-7 and the fits are presented in Figures 2,3 and 4. Tables 3,4 and 5 present
the RD, MAG and TH model fitting parameters for all events. The normal IIn, SLSN-II and SLSN-I events in each table are separated
by straight lines. Tables 6 and 7 present the CSM+RD model fitting parameters for H-rich and H-poor events respectively.
For each class of models we characterize and discuss issues of parameter
correlation and hence degeneracy that make it difficult to determine
unique parameter fits. Correlation usually increases the parameter uncertainty. 
We took this into account in our calculations and the errors reported in
the Tables reflect the parameter correlations.

\subsection{{\it Radioactive diffusion (RD) models}}

The RD model fitting and derived parameters for all events in our sample are listed
in Table 3. As can be seen, but also speculated before (Smith et al. 2007; Chatzopoulos,
Wheeler \& Vinko 2012; Gal-Yam 2012), almost no SLSN event can be powered solely by the
radioactive decays of $^{56}$Ni and $^{56}$Co due to the unphysical result of $M_{ej} < M_{Ni}$. We stress,
however, that the estimation of $M_{ej}$ using Equation 3 is uncertain given that our cited $M_{ej}$ 
values all assume $v =$~10,000~km~s$^{?1}$ and $\kappa =$~0.33~cm$^{2}$~g$^{-1}$. For the SLSN-II events this
choice for the optical opacity of the SN ejecta is reasonable (but note that prior to ionization the opacity
of the CSM is likely to be much less). For SLSN-I events a
lower opacity value of $\kappa \simeq$~0.1~cm$^{2}$~g$^{-1}$ (Valenti et al. 2008) may be more appropriate.
Such a choice would increase the $M_{ej}$ estimates for SLSN-I by a factor of 3. Even in that
case, the $M_{ej} < M_{Ni}$ inconsistency would still hold for most SLSN-I events. SN~2010gx
and SN~2010kd may be exceptions, but even their results would imply that most of the
SN ejecta are made of $^{56}$Ni, still an extraordinary condition. Some authors (Gal-Yam 2010)
have adopted an even lower opacity value ($\kappa =$~0.05~cm$^{2}$~g$^{-1}$) that leads to even higher SN
ejecta mass estimates (by a factor of 6.6) than the values shown in Table 1. The Thompson scattering
opacity in an ejecta that lacks both H and He is uncertain and will depend not
only on the chemical composition, but also on ionization conditions. It is difficult to constrain
this parameter without a detailed model of the ejecta, which is beyond the scope of this
paper. In addition to the ejecta and nickel mass constraints, it can be seen in Figures 2-4 that for many
events the very late time-decline rate is not well reproduced by RD models.

The fitted value of the LC time-scale, $t_{d}$, also has a major impact on the
estimated $M_{ej}$. As noted before, for many of the events in our sample (specifically
SN~2006tf, SN~2007bi, SN~2008iy, SN~2010kd) we lack LC data during the phase of the
rise to peak luminosity, and we therefore lack an accurate estimate of the explosion dates
of the events. For this reason, in our fitting procedure we let the explosion date (and, as a
result, the rise time to maximum light, $t_{rise}$) be a fitting parameter as well, but constrained
within a range of values suggested by the discoverers of each particular event based on
pre-explosion upper limits. Note that since the explosion
date ($t_{0}$) is a parameter that is not related to the physics of a particular SN, in Table 3 we
present the rise time ($t_{rise} = t_{max} - t_{0}$) instead. The fitted explosion dates therefore have a
direct effect on the fitted $t_{d}$ values. As we show in detail in the Appendix (Table A1), in our
fitting procedure we recover this strong correlation between $t_{d}$ and $t_{rise}$. These parameters
are often similar, but not identical.

A major impact of the explosion date uncertainty is observed in the case of the PISN
candidate SN~2007bi. Using the Gal-Yam et al. (2009) adopted value of $t_{d} =$~70~d, with
$v =$~10,000~km~s$^{-1}$ and $\kappa =$~0.1~cm$^{2}$~g$^{-1}$ appropriate for H-poor SN ejecta, we obtain
$M_{ej} =$~22.6~$M_{\odot}$ while, for $\kappa =$~0.05~cm$^{2}$~g$^{-1}$, $M_{ej} =$~45.3~$M_{\odot}$. For the lower
choice of opacity and this $t_{d}$ value, we recover the results of Gal-Yam (2009), Moriya et al.
(2010) and Yoshida \& Umeda (2011) which imply that models of PISN or energetic CCSN 
may be consistent with SN~2007bi in terms of the LC, since $M_{ej} >> M_{Ni}$.
Taking into account the uncertainty in the explosion date, however, our fitting derived a
much smaller value for $t_{d}$ (25.2~d) for which no reasonable choice for $\kappa$ satisfies the physical
$M_{ej} >M_{Ni}$ solution. This result indicates that the nature of SN~2007bi and its interpretation
as a PISN remain under debate. Given this uncertainty in the explosion date, PISN models
for SN~2007bi that involve large amounts of $^{56}$Ni are possible, at least in terms of the LC as
noted by Gal-Yam (2010). Scaling, the late, nebular spectrum of the archetypical broad-lined
Type Ic SN~1998bw associated with a GRB to match the nebular spectrum of SN~2007bi
also suggests the production of substantial $^{56}$Ni. Recent non-LTE radiation hydrodynamics models
of PISNe (Dessart et al. 2013), however, show that the spectral evolution of SN~2007bi is
inconsistent, in terms of color, temperature and spectral features, with that expected for a
PISN. We return to a discussion on alternative models for SN~2007bi in \S4.3 and \S4.4.

Our fitting tests presented in detail in the Appendix indicate that the RD model fitting
parameters are all correlated. 
In Figure A1 (see also Table A1) the red points that represent models that fit the data
in the parameter space are distributed along a line. If a fit fails to reproduce the exact 
($t_d$, $M_{Ni}$) pair, the result may be a slightly different set of values for $t_d$ and 
$M_{Ni}$ that vary from the exact solution, but still fit the data reasonably well.
The tightest correlations are observed between $t_{d}$ and $t_{rise}$,
as discussed above, but $M_{Ni}$ and $t_{d}$ are also correlated. The tight correlation
between $M_{Ni}$ and $t_{d}$ reflects ``ArnettÕs rule" (A80, 82), which is built into the
RD model by design and implies that at LC peak the input and output power are equal.
For all RD models discussed here, the gamma-ray leakage parameter, $A_{\gamma}$, is so large and
unconstrained that it is irrelevant and does not strongly affect the fitting results.

\subsection{{\it Magnetar (MAG) models}}

Table 4 lists the final fitting and derived parameters for the MAG model for all events
in our sample. The MAG model provides good, low reduced-$\chi^{2}$ fits for the majority of SN
LCs that we examine in this work. For all events, the B-field values and initial magnetar
periods implied are in ranges expected for magnetars ($B = 0.1-10 \times 10^{14}$~G, $P_{i} =$~1-4~ms) with
the exception of PTF~09uj.

The {\tt MINIM} fitting results for the MAG model also seem to be in reasonable agreement with the
results of Kasen \& Bildsten (2010) for the LCs of SN~2007bi and SN~2008es. For SN~2008es
we derive a somewhat weaker B-field and a bit slower initial magnetar period. Our derived
$M_{ej}$ using fiducial values for v and $\kappa$ is much smaller than the 5~$M_{\odot}$ presented
by Kasen \& Bildsten (2010). The discrepancy in the derived $M_{ej}$ can be attributed to
reasons similar to those discussed for the RD model (uncertain opacity and explosion time).
For SN~2007bi our agreement with Kasen \& Bildsten (2010) is much better (their fit gives
$B = 2.5 \times 10^{14}$~G, $P_{i} =$~2~ms and $M_{ej} =$~20~$M_{\odot}$; our derived $M_{ej}$ would become 24.3~$M_{\odot}$ for
the H-poor appropriate choice of $\kappa =$~0.2~cm$^{2}$~g$^{-1}$).

In the Appendix we discuss the parameter correlations for the MAG model. The MAG
model version of the ``Arnett rule" is also recovered here via strong correlations between the
$E_{p}$, $t_{p}$ and $t_{d}$ parameters. The strong anti-correlation between $t_{p}$ and $t_{d}$ is also suggested
by Kasen \& Bildsten (2010) and their Equation 16. As can be seen in Table A2, despite
the strong correlations among most parameters of the MAG model, all parameters are well
recovered in a test fitting done by {\tt MINIM}.

The MAG model is not favored for SLSN-II events for which clear signs of CSM interaction are detected in
the spectra in the form of intermediate and narrow-width emission lines. For SLSN-I events,
the MAG model cannot be ruled out, at least in terms of quality of fit to the LCs. Non-LTE
radiation hydrodynamics models recently presented by Dessart et al. (2012) suggest that the
MAG model may indeed be relevant for events such as SN~2007bi, PTF~09atu (Quimby et al.
2011) and other SLSN-I under the basic assumption that the radiation from the magnetar
thermalizes efficiently in the expanding SN ejecta (but see Bucciantini et al. 2006).

\subsection{{\it CSM-interaction (TH and CSM+RD) models}}

The most successful models for SLSN LCs in terms of reduced-$\chi^{2}$ values and physical
consistency of the derived parameters are those for which the main power source is shock
heating due to SN ejecta-CSM interaction. In our analysis here we assume that members
of both SLSN-I and SLSN-II may be powered by CSM interaction. In the case of SLSN-II
there are clear signs of such interaction in the optical spectra of these objects, specifically the
intermediate and narrow width Balmer emission lines that are formed due to recombination
that follows ionization of CSM material due to shock or radiative heating. It has been suggested (Smith
et al. 2007; 2010) that the intermediate-width emission lines are formed within the shocked
CSM and thus are related to the velocity of the forward shock, while the narrow-width components
are formed in the mediated but yet un-shocked, extended CSM material. In contrast, SN~2008es
developed only intermediate width hydrogen emission lines while narrow width components
were not detected (Gezari et al. 2009; Miller et al. 2009). The lack of narrow-width emission
could be due to a variety of reasons. One could be that the ionizing radiation from the forward
(and maybe the reverse) shock while it was in the dense shell was not strong enough to
ionize dilute material beyond the CSM shell. Another idea is that the interaction was
dominated by a fast-moving CSM shell that only gave rise to intermediate
width emission lines, and that an extended CSM was either absent or of too low density to have
an effect on the observed spectrum. Emission due to CSM
interaction may produce a blue continuum even in a hydrogen and helium deficient CSM (Gal-Yam, private communication).
The late-time (+414 d) spectrum of SN~2007bi does not show a clear
blue continuum, but at such late times it is possible that the forward shock has
already exited the optically-thick CSM shell and is propagating in CSM too dilute
to produce observable spectral features.

We thus argue that the absence of narrow line emission or well-defined blue continua in the optical spectra of SLSN
does not necessarily constitute an argument against CSM interaction. Following this line
of thought, we consider the possibility that at least some members of the SLSN-I class are
powered by H-poor CSM interaction. In such case, one might still expect the appearance of
emission lines indicative of a H-poor CSM composition in the spectra of some SLSN-I. A
possible example of such intermediate width emission arising in a H-poor CSM shell could
be the [\ion{O}{1}]~$\lambda \lambda$~6300, 6364 and \ion{O}{1}~$\lambda$~7775 features 
in the +54~d and +414~d post-maximum
spectra of SN~2007bi (Gal-Yam et al. 2009). Until detailed, non-LTE radiation
hydrodynamics models of H-poor CSM interaction models become available, the nature of
these lines as well as their formation sites (either SN ejecta or H-poor CSM) remain debatable.
Given these uncertainties, we elect to investigate models of CSM interaction even for SLSN-I
events (see Chatzopoulos \& Wheeler 2012b for a formation scenario for H-poor CSM shells).

A simplified version of CSM interaction is the TH model that assumes constant shock
energy deposition in the SN ejecta, $E_{sh}$, for a time-scale $t_{sh}$. The fitting results for the TH
models are presented in Table 5. The derived $E_{sh}$ values for all SLSNe in our sample range
from $0.5-2 \times 10^{51}$~erg, typical of SN radiated energies ($\sim$~$10^{51}$~erg) with the exception of
CSS100217 which is an outlier for all models. 
We suspect that the reason for that is the combination of high peak luminosity and very slow LC evolution
which requires both large energy input that is efficiently converted to radiation and large diffusion mass. 
In most cases, the derived $t_{sh}$ is strongly correlated with $t_{rise}$
and sets the characteristic LC time-scale and the total mass of the optically-thick SN ejecta
and CSM that is heated by the constant energy shock, $M_{CSM,th}$. Due to their unprecedented,
long duration LCs, SN~2008iy and CSS100217 imply extraordinary values for $M_{CSM,th}$. All
other events yield values that can be representative of SNe surrounded by dense shells or
optically-thick winds ($M_{CSM,th} \sim$~2-15~$M_{\odot}$ accounting for the uncertainty due to $\kappa$, especially
in the case of SLSN-I). The TH model fitting parameters are not as tightly correlated as
those of the other LC models discussed here (see Appendix).

The CSM+RD model is a more detailed version of the CSM interaction model that is
described in \S 2 and by Chatzopoulos, Wheeler \& Vinko (2012). This model also includes
contributions from the radioactive decays of $^{56}$Ni and $^{56}$Co. The CSM+RD model fitting
parameters for the H-rich events are presented in Table 6 while those for the H-poor events
are given in Table 7. In all of our fits, we fixed the slope of the inner SN ejecta density
profile to be $\delta =$~2 and that of the outer SN ejecta to be $n =$~12. We fixed the slope of the
CSM density profile to be $s =$~0, indicative of a constant density CSM shell, but we also
investigated cases of $s =$~2 characteristic of $r^{-2}$ steady-state winds to determine what type
of CSM environment is more relevant to SLSNe.

We find that all SLSNe, of both types, can be well fitted by CSM+RD models. Most
SLSN LCs are fit better under the assumption of constant density ($s =$~0), relatively massive
CSM shells. The derived $M_{CSM}$ values for SLSN-II range from $\sim$~3-5~$M_{\odot}$, with the exception
of CSS100217 which yields an extraordinary $\sim$~77~$M_{\odot}$. For SLSN-I, the derived $M_{CSM}$ range
is shifted to somewhat smaller values ($\sim$~1-4~$M_{\odot}$). This may be consistent with the less
massive H-poor shells that are expected to be ejected from more compact progenitors which
have lost their large hydrogen envelopes (Chatzopoulos \& Wheeler 2012b). The derived
SN energies and ejecta masses span the whole range between typical and energetic CCSNe
($10^{51}$~erg $< E_{SN} <$~$10^{52}$~erg; 7~$M_{\odot}$~$< M_{ej} <$~50~$M_{\odot}$). Again, CSS100217 is a significant
exception. We find typical distances of the CSM shell from the progenitor
star ($\sim$~$10^{15}$~cm). This is naturally expected because this is the radius at which radiation is most
efficiently emitted and why ordinary supernovae have photospheres at $10^{15}$~cm near maximum light.
CSM densities are characteristic of those proposed for LBV or PPISN
shells ($10^{-12}$-$10^{-10}$~g~cm$^{-3}$, Woosley, Blinnikov \& Heger 2007; Smith et al. 2007; van Marle
et al. 2010). 

The best-fit CSM+RD models also predict extreme mass-loss properties for the progenitors
of SLSNe in the years preceding the SN explosion. The derived characteristic $R_{CSM}$
of $\sim$~$10^{15}$~cm together with fiducial ejected CSM shell speeds of $\sim$~100-1,000~km~s$^{-1}$ imply
that, if CSM interaction is relevant to most SLSNe, the CSM shells associated with it were
ejected only a few months up to a few years prior to the SN explosion. A possible mechanism
could be mass-loss via gravity waves during the advanced burning stages of some massive
stars (Quataert \& Shiode 2012). Other potential mass-loss mechanisms are LBV-type mass
loss reminiscent to $\eta$-Carina (Smith et al. 2007), or shell ejection in the context of PPISN
(Woosley, Blinnikov \& Heger 2007; Chatzopoulos \& Wheeler 2012b).

We were unable to recover a good $s =$~2 fit for SN~2006gy ($\chi^{2}/dof =$~49.3),
in agreement with the results of Moriya et al. (2013). Our results for the $s =$~2 model of
PTF~09uj are in good agreement within uncertainties with those of Ofek et al. (2010) who determined
that the event could be explained as shock breakout through an optically thick wind
with characteristic wind velocity, $v_{w} =$~100~km~s$^{-1}$ and mass-loss rate, $\dot{M} =$~0.1~$M_{\odot}$~yr$^{-1}$.
Using our best fitted $\rho_{CSM,1}$ and $R_{p}$ values we estimate $\dot{M} =$~0.33~$M_{\odot}$~yr$^{-1}$ for the same
value for $v_{w}$. The SLSN-I SN2010gx is also best fit by a $s =$~2 CSM+RD model that implies
$\dot{M} =$~1.33~$M_{\odot}$~yr$^{-1}$ for $v_{w} =$~100~km~s$^{-1}$ ($\dot{M}$ can be scaled down for higher choices for $v_{w}$).
That result is in agreement with the findings of Ginzburg \& Balberg (2012) that a dense
CSM wind with $\rho \sim r^{-2}$ can reproduce the LC.

Although the CSM+RD model provides the best and most physically consistent fits to
the LCs of SLSNe, it can be argued that this is due to the large number of fitting parameters
associated with it. We argue, regardless, that it is relevant to attempt to fit SN LC CSM+RD
models because they also capture some of the natural complexity of the CSM interaction
phenomenon. We find most of the CSM+RD parameters to be tightly correlated or anticorrelated
with each other, with the natural exception of $M_{Ni}$. In addition we find that for
all best fit CSM+RD models the additional RD input has a negligible effect on the final
output LC even though in some cases large $M_{Ni}$ masses are found.

\section{SUMMARY AND CONCLUSIONS}

Making use of the capabilities of the new $\chi^{2}$-minimization code {\tt MINIM} we fit three
main semi-analytic SN LC input power models (RD, MAG, CSM+RD) to the observed
pseudo-bolometric LCs of a sample of 10 SLSNe (5 SLSN-I and 5 SLSN-II events) and 2
normal luminosity SNe IIn. Our fitting procedure with {\tt MINIM} allowed for the calculation
of fitting parameter uncertainties, correlation and degeneracy which helped us assess the
quality and physical implications of our fits. We then evaluated our results and determined
which models best fit each individual event taking into account the consistency with other
observations such as spectroscopy. The basic aim of this study was to provide insight to the
question of the observed diversity of SLSN LCs and to understand if it arises as a result of
one dominant mechanism that involves many parameters or by a combination of different
power mechanisms. We also test the power of the semi-analytic CSM+RD model that we
introduced in an earlier work (Chatzopoulos, Wheeler \& Vinko 2012) by applying it to the
LCs of normal and superluminous Type IIn events. The main results of our study are summarized
below:

\begin{itemize}
\item{The derived black-body temperatures of SLSNe at peak luminosity are similar
between the two spectroscopic types. Specifically, SLSN-I do not appear to be
systematically hotter than SLSN-II.}
\item{The semi-analytic CSM+RD model can reproduce the available numerical model LCs
for SN~2006gy, accounting for the model assumptions and uncertainties.}
\item{The power of the semi-analytic CSM+RD model is that it can be easily fit to observed
SN LCs and be then used as a reference point for more detailed radiation hydrodynamics
calculations.}
\item{The lack of observations during the rising part of the LC for a few SLSNe poses a serious
problem in determining accurate explosion dates that, in turn, introduces large
uncertainties in estimates of ejecta mass. The LCs of SN~2006tf and SN~2010kd
but maybe also SN~2007bi, are examples of this issue.}
\item{The LCs of the majority of SLSNe cannot be powered solely by the radioactive decays
of $^{56}$Ni and $^{56}$Co (RD model) because the $^{56}$Ni mass needed to power their peak luminosities
either exceeds or is close to the total SN ejecta mass implied from the duration
of the LC, for a reasonable choice of SN ejecta opacities.}
\item{Models of magnetar spin-down (MAG model) provide reasonable fits to the LCs of
most SLSNe, but are more relevant to SLSN-I. A significant uncertainty for the MAG
models is the issue of the thermalization of pulsar radiation in the expanding SN
ejecta.}
\item{CSM interaction models provide the best fits to all SN LCs in our sample. These
models are certainly relevant to the SLSN-II category where clear signs of H-rich CSM
interaction are seen in the spectra and cannot be ruled out for SLSN-I events.}
\item{In most cases, models of constant density CSM shells ($s =$~0) provide better fits than steady-state
winds ($s =$~2) to the LCs of SLSNe. That could mean that the environments around
extreme SNe are also extreme, possibly formed via episodic mass-loss and shell ejection
events.}
\item{The CSM interaction models imply SN ejecta masses (7-50~$M_{\odot}$), SN energies 
($1-10 \times 10^{51}$~erg), and CSM masses (1-5~$M_{\odot}$) that are appropriate for high-mass progenitor stars.}
\item{The CSM shells around SLSN-I are found to be somewhat less massive than those
around SLSN-II.}
\item{For SN~2007bi a hybrid model of
H-poor CSM interaction plus radioactive decay model in which the bulk of the energy
is supplied by the interaction provides a decent LC.
The lack of narrow emission lines and distinct late-time blue continuum do not necessarily
constitute lack of H-poor CSM interaction.}
\item{The extreme CSM environments and mass-loss rates implied by the CSM interaction
models indicate that the progenitors of these events were probably
quite massive and exploded via energetic CCSNe. With the exception of CSS100217 the combined $M_{ej}$
and $M_{CSM}$ imply progenitor masses significantly smaller than the mass limits for
PISNe. The mass loss mechanisms for
these progenitors remain unknown; however LBVs, PPISNe and mass-loss via gravity
waves are some potential candidates.}
\item{For all LC models investigated in this work most of the fitting parameters are found
to be tightly correlated with each other, and hence strongly degenerate.
This usually increases the uncertainties of the best-fit parameters and
may cause systematic deviations from the true values, especially in
the CSM+RD model which has the largest number of parameters. Thus,
despite of the success of the LC models in reproducing the results
from numerical simulations, one should interpret the best-fit parameters
with caution.}
\item{The diversity of SLSNe in terms of LCs and composition could be
the natural result of the diversity of CSM environments around massive progenitor
stars, including CS material that is both H and He deficient.}
\end{itemize}

There is growing evidence for hydrogen--deficient circumstellar matter. There are a number of
events categorized as Type Ibn, by dint of having no evidence of hydrogen, but narrow
emission lines of helium corresponding to a photoionized, slowly--moving CSM 
(Pastorello et al. 2008). The SN~Ibn are clearly He rich, but there may be some
configuration in which helium is present but not so strongly excited so that it
more difficult to detect directly. Another alternative is that the CSM is deficient
in both hydrogen and helium (Chatzopoulos \& Wheeler 2012b). Possible 
examples of this are SN~2006oz (Leloudas et al. (2012) and the Pan-STARRS
discoveries PS1-10ky and PS1-10awh (Chomiuk et al. 2011), PS1-11bam (Berger
et al. 2012), PS1-10afx (Chornock et al. 2013) and PS1-10bzj (Lunnan et al. 2013).
In this case,
one might expect lines of carbon or oxygen of intermediate or narrow width, but
again such line emission may depend on the distribution, motion, and ionization
of the CSM. If there is a single physical mechanism that accounts for all SLSNe, then
economy of hypotheses argues that it is CSM interaction as strongly
indicated for the SLSN-II.
Our main result is that SLSN LCs of both types can be reasonably reproduced by
CSM+RD models. This raises the importance of accurately modeling the radiation from
CSM interaction that involves a variety of geometries and compositions for the
CSM. 

The suggestion that CSM interaction is the common process
could have an impact on the interpretation of some SLSN-I events,
especially SN~2007bi, as PISNe. Kasen et al. (2011) explored PISN model spectra
for SN~2007bi, however Dessart et al. (2013) noted that the model spectrum
that Kasen et al. compared to the observed one at +51~d after explosion was for a
much earlier phase (by $\sim$~100 d). At later times the model PISN spectra from
Kasen et al. will no longer be sufficiently blue, in contradiction with the
observations of SN~2007bi.
SN~2010kd had spectra rather similar to SN~2007bi (Vinko et al. 2013, in preparation), but it is very difficult to fit the
LC with an RD model because the $^{56}$Ni mass must be comparable to or exceed the ejecta mass. This result
also casts additional doubt that SN~2007bi must necessarly be a PISN.

If events like SN~2007bi are not PISNe, the most likely alternative is that they
result from energetic CCSNe
with very massive progenitors. This conclusion is in agreement with the models of Moriya et al.
(2010) and Yoshida \& Umeda (2011) for SN~2007bi, even though in our models the majority
of the input energy is provided by CSM interaction instead of the radioactive
decays of $^{56}$Ni and $^{56}$Co. 

One aspect of the CSM+RD model is that
by its nature it poses a difficulty in definitively unveiling the characteristics of the
progenitor stars. Most of the radiation is produced by CSM interaction and
diffused within an optically-thick CSM shell that obscures the SN ejecta, at least
at early times. In our CSM+RD parameter study,
we find that $R_{0}$ and $M_{ej}$, both parameters relevant to the SN progenitor,
are the most weakly constrained.

The large parameter space associated with the CSM+RD model is the natural
consequence of the large diversity physically associated with CSM interaction;
a variety of combinations of progenitor ($v_{SN}$, $R_{p}$, $M_{CSM}$ and $n$) and
CSM ($M_{CSM}$, $s$, $\rho_{CSM,1}$) characteristics can yield a variety of LC shapes, durations,
peak luminosities and decline rates. Different sets of parameters may also yield very similar LCs 
because of parameter degeneracy.
In reality the potential diversity is even larger than implied by the parameters of
the CSM+RD model we discuss here; effects of different CSM geometry (bipolar
shells, circumstellar disks (Metzger et al. 2010) or clumps (Agnoletto et al.
2009) and composition (H-rich vs H-poor and/or metal-rich) must also play a
role. In some ways, just looking at the famous Hubble image of $\eta$-Carina is itself
an illustration of the complexity of CSM environments that can exist around
massive evolved stars.

As emphasized here, the extraordinary properties of SLSNe will be probed
at a more profound level only via accurate non-LTE radiation hydrodynamical
modeling for all different power input mechanisms that will allow for direct comparison
not only with observed LCs but also spectra of contemporaneous phases.
As recently indicated by the findings of Dessart et al. (2012, 2013), reproducing
SLSN LCs alone does not constitute a definitive answer about the nature of a
particular event. Understanding of mass loss mechanisms during
the late stages of massive stellar evolution will help unveil how extreme CSM
environments are formed around SLSN progenitors and the exact role they play
in giving rise to the observed radiative properties of individual SLSNe.

We would like to thank the anonymous referee,
Roger Chevalier, Vikram Dwarkadas, Sean Couch, Todd Thompson
and Takashi Moriya for useful discussions and comments. This research is supported by
NSF Grant AST 11-09801 to JCW. EC would like to thank the University of Texas Graduate
School William C. Powers fellowship for its support of his studies. JV is supported by
Hungarian OTKA Grants K76816 and NN107637.


{}                     

\appendix
\section{APPENDIX: Fit parameters and parameter correlations}

In this Appendix we give some details on the fitting of the free parameters of each
model and discuss the correlation between parameters as measured by their covariance and
correlation matrices. To do so, we first synthesized a generic test LC for each model using
the equations given in \S2. We then ran the code {\tt MINIM} to independently determine the free
parameters and to determine the correlations among them. This also served as a test for
the reliability of {\tt MINIM} and the optimization algorithm we applied (see \S 3). The fitting was
successful in all cases, since the initial parameters were recovered within the uncertainties.

\subsection{Radioactive diffusion (RD) model.}

First introduced by A80, A82, this model assumes a spherical, homologously expanding
ejecta. The energy input generated by the decay of radioactive $^{56}$Ni and $^{56}$Co slowly diffuses
out from the center to the surface. The resulting LC is expressed as Equation 1 in \S2, where
the meaning of the symbols are also explained. We take advantage of having $v t_{d} >> R_{0}$
for the SNe we consider, thus, the terms involving $R_{0}/v t_{d}$ can be ignored with respect to
$t/t_{d}$. This considerably reduces the number of free parameters in this model, resulting in the
following:
\begin{itemize}
\item{$t_{ini}$: the initial epoch of explosion, expressed in days relative to a pre-selected fiducial
explosion time $t_{exp}$ (see Table 2) for each SN},
\item{$M_{Ni}$: the initial mass of radioactive $^{56}$Ni (in $M_{\odot}$) synthesized in the explosion,}
\item{$t_d$: the effective LC time-scale (in days), sometimes termed as diffusion timescale by several
authors,}
\item{$A_{\gamma}$: optical depth of the SN ejecta to gamma-rays, measured at +10 days after explosion.}
\end{itemize}

Figure A1 shows the distribution of random choices of free parameters around the $\chi^{2}$
minimum for four particular parameter combinations. The general shape of this distribution
illustrates the correlation between the two particular parameters: a nearly symmetric
distribution means less correlation (parameters are independent), while an elongated shape
indicates that these parameters are correlated. In the latter case, if the two parameters are
slightly changed according to the direction indicated by the curvature of the $\chi^{2}$ hypersurface,
the output LC remains almost the same. In other words, these parameters cannot be fully
recovered in every case; only their linear combination can be determined by the fitting of
the LC.
Figure A1 suggests that none of the parameters of the RD model are independent (as
also implied by the physics of the model), they are more-or-less correlated with each other.
We estimated the correlation between them for each parameter combination by calculating
\begin{equation}
R_{i,j} = \frac{{\sum_{k=1}^{N_r} (p_i(k) - p_i(min) \cdot (p_j(k) - p_j(min)) }}{{(N_p -1) \sigma_i \sigma_j }},
\end{equation}      
where $N_r = 200$ is the number of random vectors (parameter sets) used in {\tt MINIM}, 
$N_{p}$ is the number of free parameters in the particular model,
$p_i(k)$ is the $i$th parameter in the $k$th vector, $p_i(min)$ is the $i$th parameter
in the parameter vector corresponding to the minimum of the $\chi^2$, and $\sigma_{i}$ is the
standard deviation of the $i$th parameter around $p_i(min)$. The correlation
coefficients can be found in Table~A1. 

Both Figure A1 and Table~A1 show that the correlation between each
pair of the physical parameters ($M_{Ni}$, $t_d$ and $A_\gamma$) is stronger
than 50 \%. The $\sim 90$ \% correlation between $M_{Ni}$ and $t_d$ is known as 
the ``Arnett-rule": at LC peak the input and output power is the same, for example,
if the peak occurs later ($t_d$ is longer), then a given peak luminosity needs more initial 
$M_{Ni}$. 

\subsection{Magnetar (MAG) model.}

Equation 2 describes the resulting LC when the power input is due to the spin-down of
a rapidly rotating magnetar (magnetized neutron star) in the center of the SN ejecta. 
As explained in \S 2, we have optimized the following LC parameters:
\begin{itemize}
\item{$t_{ini}$: the initial epoch of explosion (in days)}
\item{$R_{0}$: the radius of the progenitor (in $10^{13}$ cm)}
\item{$E_{p}$: the initial rotational energy of the magnetar (in $10^{51}$ erg)}
\item{$t_{d}$: the effective LC time-scale, as in the RD model (in days)}
\item{$t_{p}$: the spin-down timescale of the magnetar (in days)}
\item{$v$: the expansion velocity of the SN ejecta (in $10^3$ km s$^{-1}$).}
\end{itemize}

The model parameters applied for a test LC and their correlation coefficients
are collected in Table~A2. Figure A2 displays the
distribution of the random vectors in the vicinity of the $\chi^2$ minimum.

As expected, $R_{0}$ and $v$ are only weakly constrained parameters, since the LC is not
sensitive to their combination of $R_{0} / v t_{d}$ appearing in Equation 2, the same as in the case 
of the RD model. All other parameters could be well recovered, despite the strong
correlations between $E_{p}$, $t_{p}$ and $t_{d}$. 

\subsection{CSM shell with top-hat energy input (TH) model.}

The third model consists of the simple configuration of a thick CSM shell around the SN
in which the power input is constant for a certain amount of time then it switches off. The
observed LC is governed by the diffusion of thermalized photons to the photosphere that is
fixed at the outer radius of the shell. This toy model has the following free parameters:
\begin{itemize}
\item{$t_{ini}$: the initial epoch of explosion (in days)}
\item{$E_{sh}$: the total input energy (in $10^{51}$ erg)}
\item{$t_{sh}$: the time interval for the constant energy input (in days)}
\item{$t_{0}$: the diffusion time in the CSM shell having a fixed photosphere.}
\end{itemize}

Table~A3 lists all the parameters of the test model and their
correlation coefficients, while in Figure A3 the $\chi^2$ function 
around the minimum is mapped. 

This model has the advantage of having relatively few parameters, and they are less strongly
correlated than those of the other models. 
More specifically, $t_{sh}$ seems to be correlated with $t_{rise}$ but not $t_{d}$.
Its drawback is, of course, the less 
physical reality of its assumptions. 

\subsection{Shock-heated ejecta and CSM-collision combined with radioactive heating (CSM+RD) model.}

This is the most complicated model, where the output LC can be calculated from Equations 4, 5 and 6
(see \S 2 for the details). The large number of parameters make this model rather ill-constrained.
In order to keep the model manageable, we have restricted the number of free parameters to 7:
\begin{itemize}
\item{$t_{ini}$}: as before (in days)
\item{$R_p$: radius of the SN progenitor prior to explosion}
\item{$M_{ej}$: ejecta mass (in $M_\odot$)}
\item{$M_{CSM}$: total mass of the CSM (in $M_\odot$)}
\item{$\rho_{CSM,1}$: density of the CSM at radius $R =$~$R_{p}$ (in $10^{-12}$~g~cm$^{-3}$)}
\item{$M_{Ni}$: initial mass of $^{56}$Ni (in $M_\odot$) }
\item{$v_{SN}$: maximum expansion velocity of the SN ejecta (in $10^3$ km s$^{-1}$).}
\end{itemize}

We elected to use $v_{SN}$ as a free parameter describing the SN kinetic energy via 
$E_{SN} = 3/10 \times (n?3)/(n? 5) \times M_{ej}(x_{0}v_{SN})^{2}$, where $n$ is the ejecta density slope parameter and $x_{0}$ is
the fractional radius of the core in the SN ejecta (see \S 2).

The remaining parameters were kept fixed to their pre-selected fiducial value: we have
applied $\kappa =$~0.33~cm$^{2}$~g$^{-1}$, n = 12, $\delta =$~2, $\beta =$~13.8, $x_{0} =$~0.1, 
and $s =$~0, 2 (see \S2 for explanation).
Our tests showed that the LC is not particularly sensitive to these parameters, except for
the CSM density slope parameter, $s$, where the $s =$~0 (constant CSM density) and the $s =$~2
(stellar wind with constant mass-loss rate) assumptions resulted in quite different LCs.

As a test case we have computed a model by assuming $s =$~0, which was then refitted
using {\tt MINIM} with both $s =$~0 and $s =$~2 ($s$ was kept fixed during the fitting). Figure A4
shows the original model LC (dots) and the fit results (solid and dotted lines). It is seen
that the shape of the LC can be recovered quite well. The original and recovered parameters
(Table A4; Figure A5), however, reveal larger differences than for the previous three models.

Given the larger number of free parameters, and the complicated nature of this model,
it is not surprising that the correlation between most parameters is quite high. Moreover,
there is a general ambiguity related to the choice of the CSM density parameter $s$. As
seen in Table A4 and Figure A5, the resulting fit parameters assuming $s =$~2 can be quite
off from the original ones computed assuming $s =$~0. $R_{0}$ and $M_{ej}$ are the most weakly
constrained parameters, as there are order of magnitude differences between their original
and reconstructed values for $s =$~2. The other parameters can be recovered within a factor
of 2-3. Higher uncertainties for $R_{0}$ and $M_{ej}$ are also seen even using the original value of
$s =$~0 during the fitting. Thus, it is concluded that even though the shape of the LC can be
relatively well described by the CSM+RD model either assuming constant density ($s =$~0)
or wind-like ($s =$~2) CSM structure, the resulting fit parameters for the SN ejecta ($R_{0}$ and
$M_{ej}$) may be off by an order-of-magnitude from their real values because of the incorrect
assumed value of $s$. The CSM-related parameters ($M_{CSM}$ and $\rho_{CSM,1}$) might be recovered
with slightly better accuracy, but those are still only weakly constrained.

\begin{figure}
\begin{center}
\includegraphics[angle=-90,width=16cm]{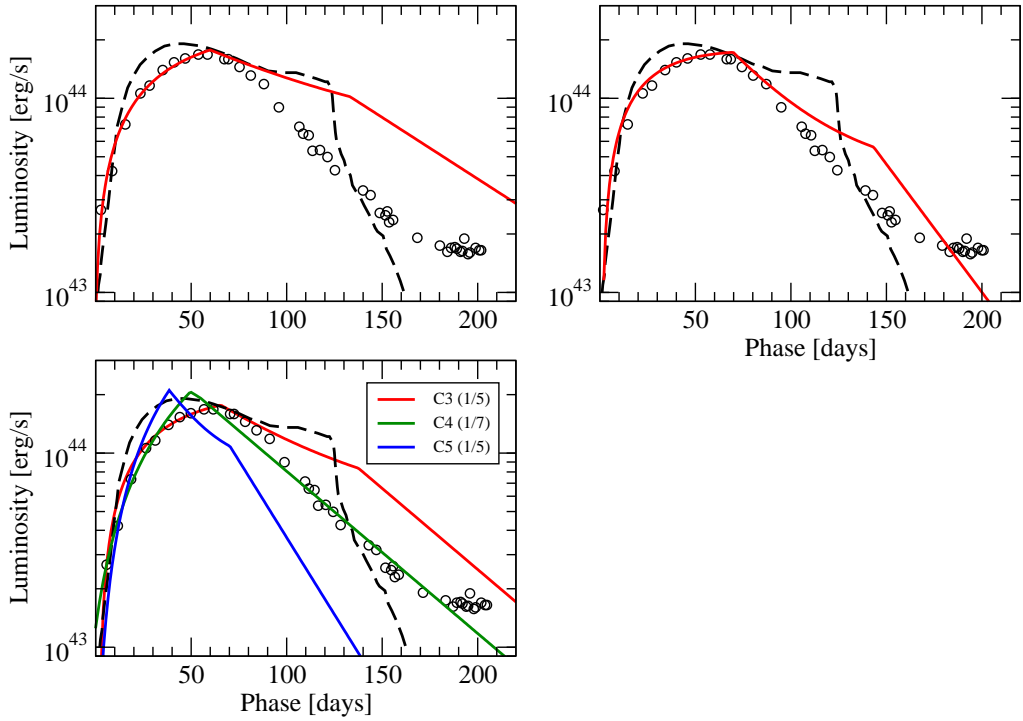}
\caption{Comparison of the LCs from the hybrid CSM+RD model with the
observed KAIT photometry of SN~2006gy. The model parameters are listed
in Table 1. The dashed curve shows the model LC computed with the
same parameters as model F1 of Moriya et al. (2013) scaled to fit
the observed maximum luminosity. Our additional models are plotted
as colored continuous curves: model C1 (upper left panel), model
C2 (upper right panel) and models C3, C4 and C5 (lower left panel;
the models are scaled to the observations with a factor indicated
in the legends). Model C4 provides a reasonable fit to the observations.}
\end{center}
\end{figure}

\begin{figure}
\begin{center}
\includegraphics[angle=-90,width=16cm]{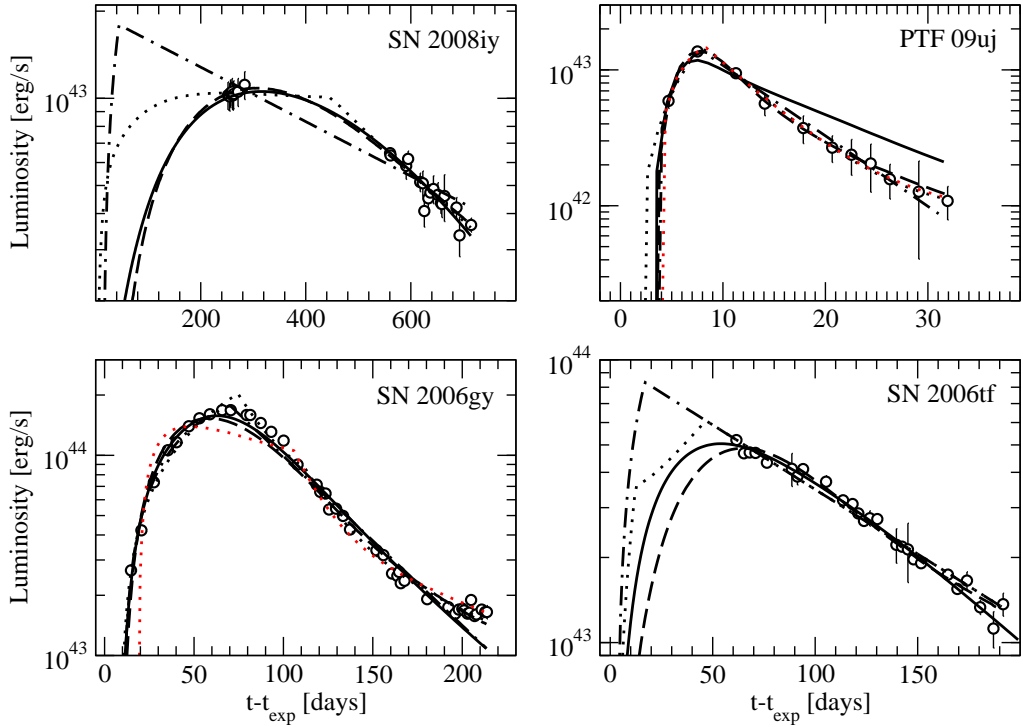}
\caption{Lowest $\chi^{2}$ model fits to the observed LCs of SN~2008iy (upper left panel), PTF~09uj (upper right panel),
SN~2006gy (lower left panel) and SN~2006tf (lower right panel). Solid curves correspond to the RD model, 
dashed curves to the MAG model, dashed-dotted curves to the TH model
and dotted curves to the hybrid CSM+RD model in the case of constant density CSM shell ($s =$~0).
The red dotted curves show the best-fit CSM+RD
model for the choice $s =$~2 (steady-state wind CSM);
this model is inferior to all the others for SN~2006gy. Parameters of
the best fit models are given in Tables 3 through 7.}
\end{center}
\end{figure}

\begin{figure}
\begin{center}
\includegraphics[angle=-90,width=16cm]{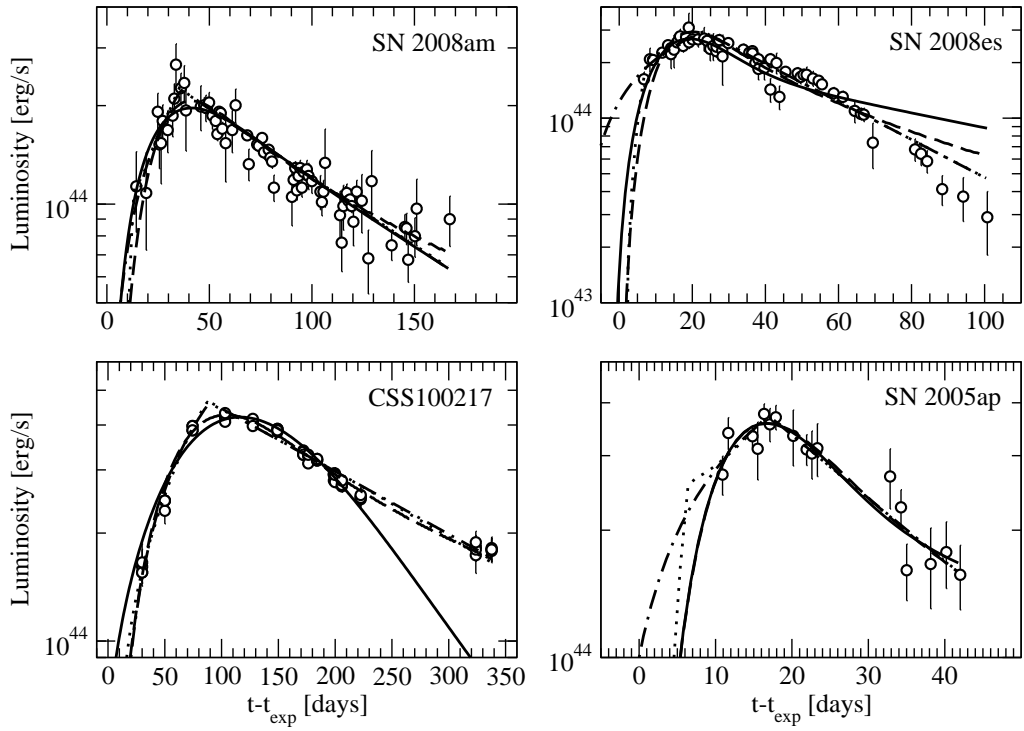}
\caption{Same as Figure 2 but for SN~2008am (upper left panel), SN~2008es (upper right panel),
CSS100217 (lower left panel) and SN~2005ap (lower right panel).}
\end{center}
\end{figure}

\begin{figure}
\begin{center}
\includegraphics[angle=-90,width=16cm]{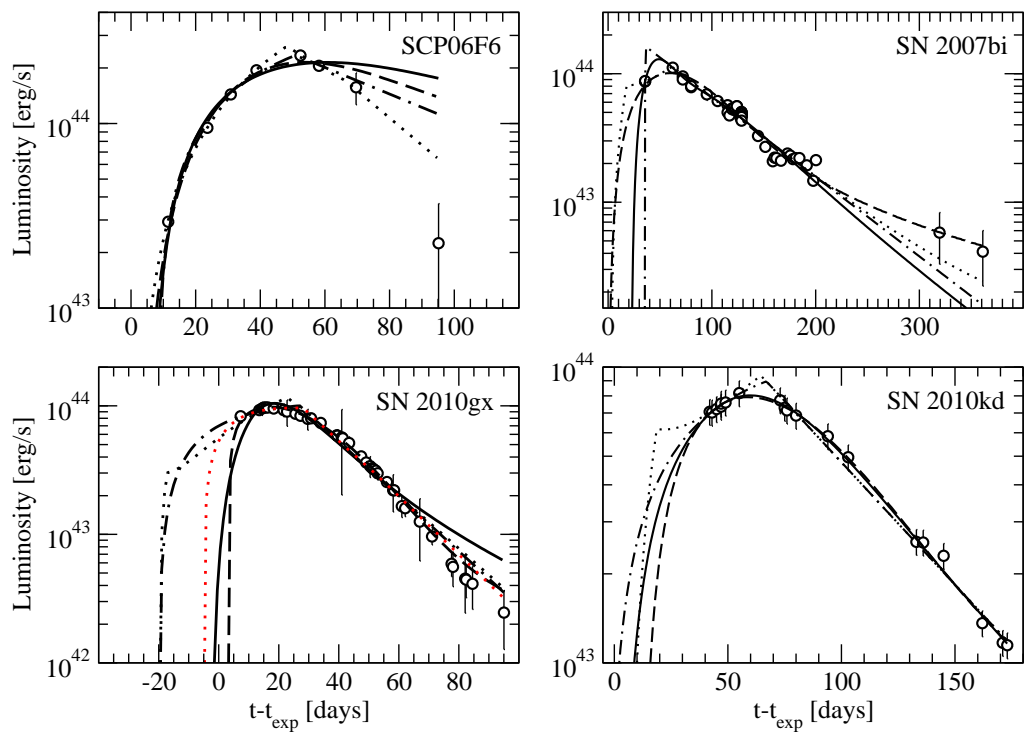}
\caption{Same as Figure 2 but for SCP06F6 (upper left panel), SN~2007bi (upper right panel),
SN~2010gx (lower left panel) and SN~2010kd (lower right panel).
Negative $t-t_{exp}$ implies that the model has an
earlier date for the explosion than the pre-selected fiducial $t_{exp}$ value
given in Table 2.}
\end{center}
\end{figure}

\begin{figure}
\begin{center}
\includegraphics[width=16cm]{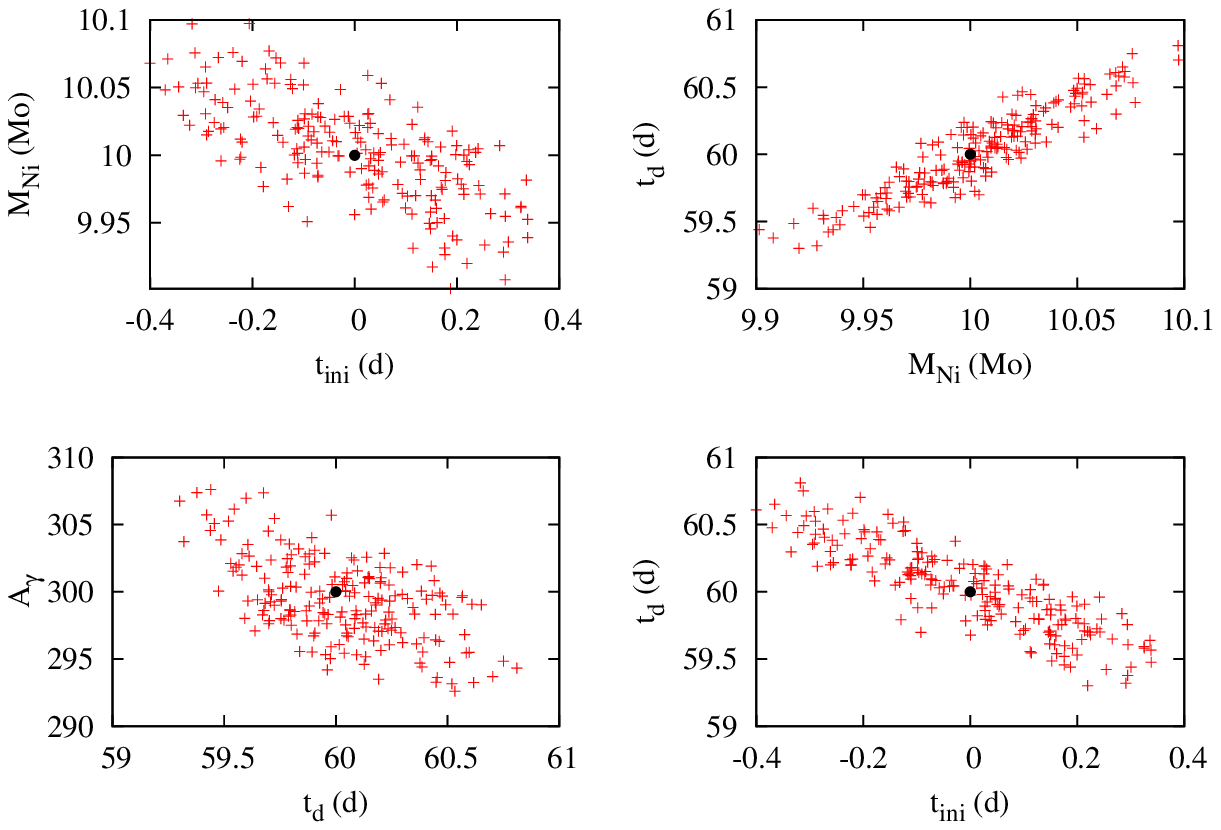}
\renewcommand\thefigure{A1} 
\caption{Distribution of random choices of parameters around the $\chi^2$ minimum found by {\tt MINIM}
with respect to a generic specified RD model. 
The extension of the distribution is $\Delta \chi^2 = 1$ corresponding to the
67 \% confidence interval around the minimum (see \S 3). More elongated ellipsoids 
indicate stronger correlation between the parameters. A filled circle shows the position 
of the initial model. For the definition of the fitting parameters plotted here please see \S A1.}
\label{fig-app1}
\end{center}
\end{figure}

\begin{figure}
\begin{center}
\includegraphics[width=16cm]{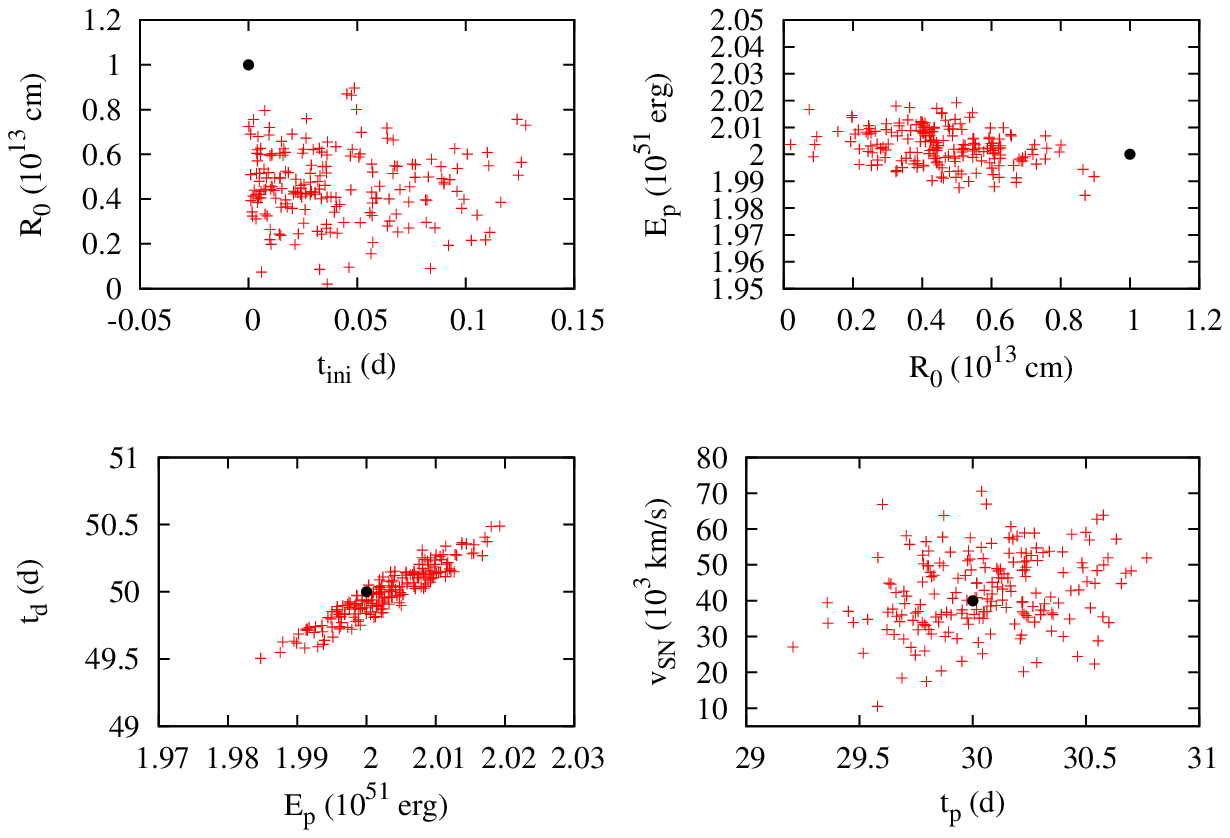}
\renewcommand\thefigure{A2} 
\caption{The same as in Figure A1, but for the MAG model. Note that $R_{0}$ is especially ill-constrained.
For the definition of the fitting parameters plotted here please see \S A2.}
\label{fig-app2}
\end{center}
\end{figure}

\begin{figure}
\begin{center}
\includegraphics[width=16cm]{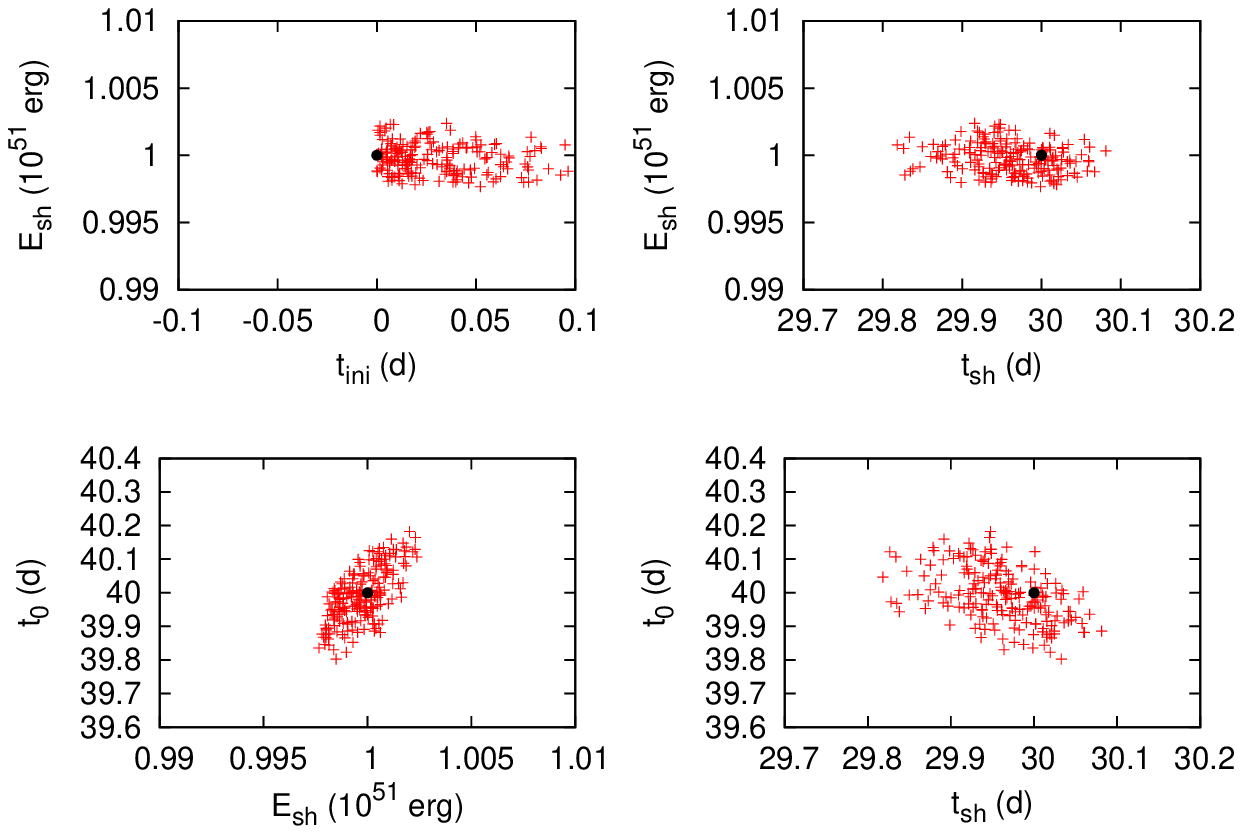}
\renewcommand\thefigure{A3} 
\caption{The same as in Figure A1, but for the TH model.
For the definition of the fitting parameters plotted here please see \S A3.}
\label{fig-app3}
\end{center}
\end{figure}
 
\begin{figure}
\begin{center}
\includegraphics[width=16cm]{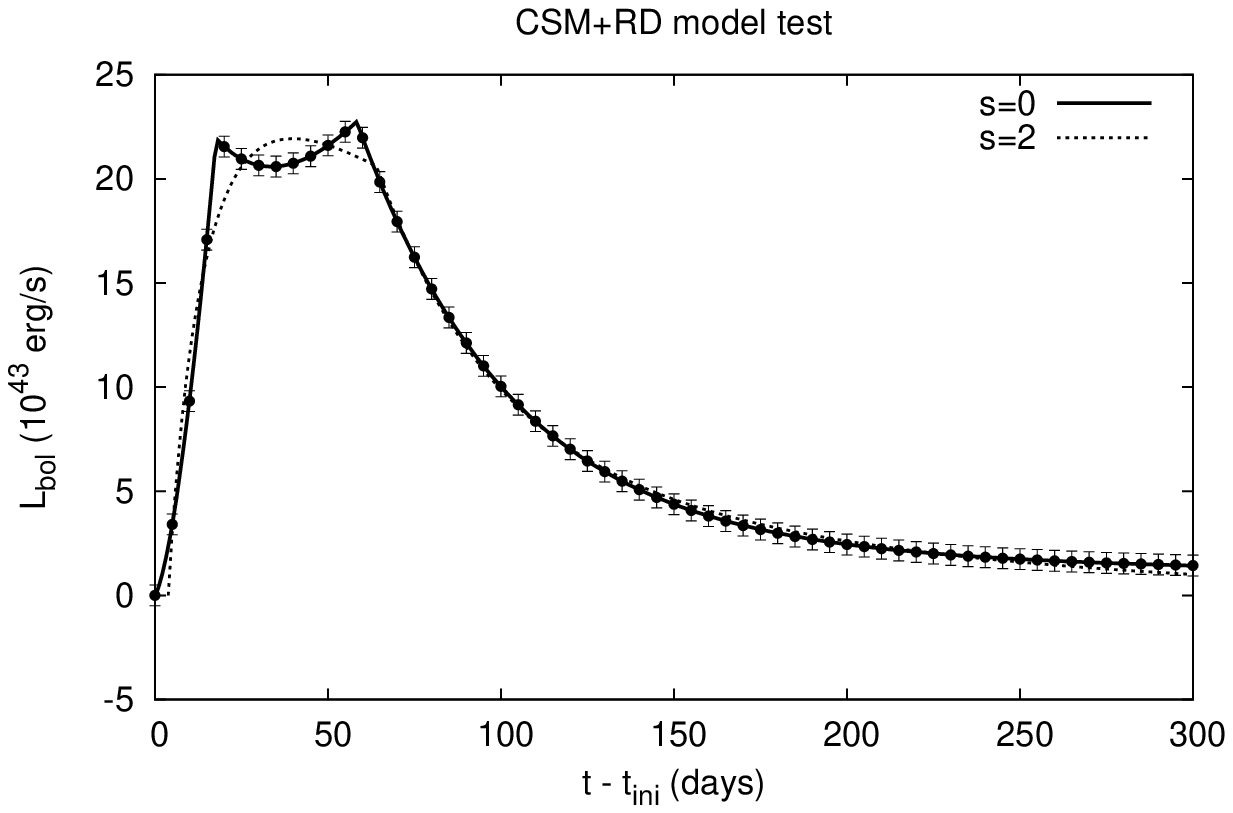}
\renewcommand\thefigure{A4} 
\caption{Comparison of CSM+RD models having different density exponents $s=0$
(constant density CSM)
and $s=2$ (wind-like CSM). The dotted curve represents the initial synthetic model
computed assuming
$s=0$ (see Table A5). The solid curve shows the best-fit $s=0$ model found by
{\tt MINIM}.
As seen in Table A5, the parameters of the initial model are recovered very well.
Dotted curve shows the best-fit $s=2$ model. Although the shape of the LCs are
similar,
the parameters are quite different in the latter case (Table A5).}
\label{fig-app4}
\end{center}
\end{figure}

\begin{figure}
\begin{center}
\includegraphics[width=16cm]{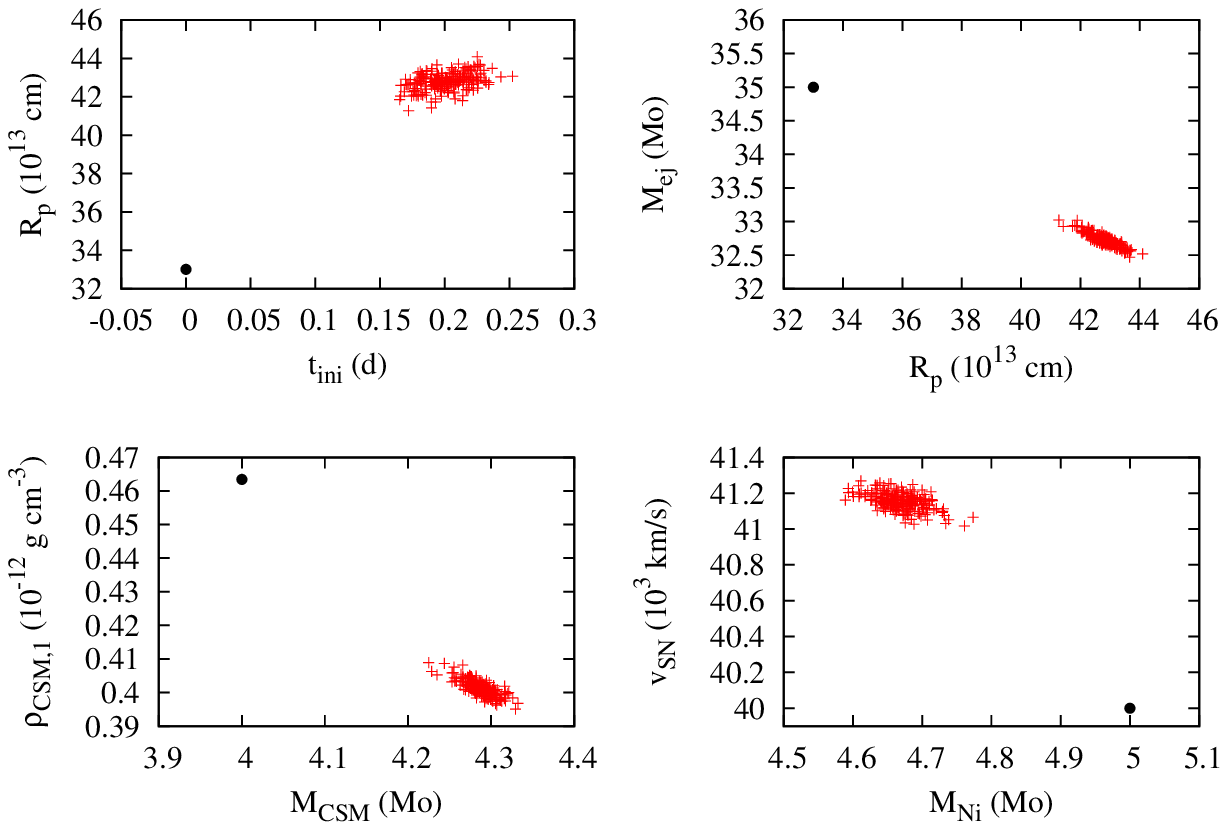}
\renewcommand\thefigure{A5} 
\caption{The same as in Figure A1, but for the CSM+RD model assuming $s =$~0 for both the synthetic light curve and the fit model.
For the definition of the fitting parameters plotted here please see \S A4.}
\label{fig-app5}
\end{center}
\end{figure}

\clearpage
\begin{deluxetable}{lcccccccc}
\tabletypesize{\tiny\tiny\tiny}
\tablewidth{0pt}
\tablecaption{Summary of the parameters of the CSM+RD models presented in Figure 1 and compared to the F1 model of Moriya et al. (2013).}
\tablehead{
\colhead {Parameter} &
\colhead {C1} &
\colhead {C2} &
\colhead {C3} &
\colhead {C4} &
\colhead {C5} &
\colhead {F1} &\\}
\startdata
$s$                          &  0   & 0    & 0    &  0   & 0	& 0    \\
$\kappa$ (cm$^2$~g${-1}$)    &  0.2 & 0.09 & 0.33 & 0.33 & 0.33 & 0.34 \\
$R_{CSM}$ ($10^{15}$~cm)     &  2.4 & 2.0  & 6.0  & 6.0  & 6.0  & 6.0  \\
$M_{CSM}$ ($M_\odot$)        &  15  & 15   & 15   & 15   & 5.0  & 15   \\
$n$                          &  7   & 7    & 7    & 12   & 12	& 7    \\
$E_{SN}$ ($10^{51}$~erg)     &  2.2 & 1.7  & 10   & 10   & 10	& 10   \\
$M_{ej}$ ($M_\odot$)         &  20  & 20   & 20   & 20   & 20	& 20   \\
$R_p$ ($10^{13}$~cm)         &  1.0 & 0.9  & 10   & 10   & 10	&  -   \\
\enddata 
\end{deluxetable}

\setcounter{table}{1}
\begin{deluxetable}{lccccccc}
\tabletypesize{\tiny\tiny\tiny}
\tablewidth{0pt}
\tablecaption{Basic data for the studied SLSNe.}
\tablehead{
\colhead {SN} &
\colhead {Type} &
\colhead {$z$} &
\colhead {$t_{exp}$~(MJD)} &
\colhead {$L_{peak}$~($10^{43}$~erg~s$^{-1}$)} &
\colhead {$T_{BB}$~($10^{4}$~K)} &
\colhead {Reference} &\\}
\startdata
SN2008iy     & IIn	 &  0.041   & 54356       & 1.12    & 1.6   & Miller et al. (2010)	\\
PTF09uj      & IIn	 &  0.065   & 55000 	  & 1.41    & 1.7   & Ofek et al. (2010)  	\\
\hline
SN2006gy     & SLSN-II   &  0.019   & 53967 	  & 21.40   & 1.2   & Smith et al. (2007)        \\
SN2006tf     & SLSN-II   &  0.074   & 54050 	  & 5.20    & 0.8   & Smith et al. (2008)        \\
SN2008am     & SLSN-II   &  0.234   & 54439 	  & 26.73   & 1.2   & Chatzopoulos et al. (2011) \\
SN2008es     & SLSN-II   &  0.202   & 54574 	  & 31.04   & 1.4   & Gezari et al. (2009)       \\
CSS100217    & SLSN-II 	 &  0.147   & 55160	  & 42.02   & 1.6   & Drake et al. (2011)        \\
\hline		
SN2005ap     & SLSN-I    &  0.283   & 53430	  & 37.02   & 2.0   & Quimby et al. (2007)       \\
SCP06F6      & SLSN-I    &  1.189   & 53772	  & 23.72  & 1.4   & Barbary et al. (2009)$^{a}$\\
SN2007bi     & SLSN-I    &  0.129   & 54089	  & 11.10   & 1.2   & Gal-Yam et al. (2009)      \\
SN2010gx     & SLSN-I  	 &  0.230   & 55260	  & 9.71    & 1.5   & Pastorello et al. (2010)   \\
SN2010kd     & SLSN-I    &  0.101   & 55483	  & 8.20    & 1.4   & Vinko et al. (2012)        \\    
\enddata 
\end{deluxetable}

\setcounter{table}{2}
\begin{deluxetable}{lcccccc}
\tabletypesize{\tiny\tiny\tiny}
\tablewidth{0pt}
\tablecaption{Best-fit parameters for the RD model.}
\tablehead{
\colhead {SN} &
\colhead {$t_{rise}$~(d)} &
\colhead {$M_{Ni}$~($M_{\odot}$)} &
\colhead {$t_{d}$~(d)} &
\colhead {$\chi^{2}/dof$$^{a}$} &
\colhead {$M_{ej}^{b}$~($M_{\odot}$)} &\\}
\startdata
SN2008iy		&315.00 (2.38)	&13.18 (0.13) & 509.30 (4.49) & 0.44 & 363.60 \\
PTF09uj      	&5.00 (0.06)	& 0.234 (0.002) & 3.00 (0.02)  & 5.63 & 0.01 \\
\hline
SN2006gy    	&56.00 (0.08) 	& 22.76 (0.24)	& 56.81 (0.43) & 13.76 & 4.50 \\
SN2006tf     	&53.88 (0.50)	& 5.94 (0.03) 	& 46.47 (0.52) & 1.94   & 3.00\\
SN2008am  	&44.00 (7.85)	& 20.45 (13.82) & 39.11 (32.53) & 1.51  & 2.14 \\
SN2008es   	&22.00 (2.36) 	& 16.07 (0.60) & 19.79 (1.98)	& 4.96 &	0.55\\
CSS100217	& 115.40 (5.60)	& 104.60 (4.70)& 127.50 (7.81) & 6.53 & 22.78 \\
\hline	
SN2005ap   	& 16.00 (0.40) 	& 15.98 (0.33) & 13.68 (0.42) & 0.60 & 0.26 \\
SCP06F6     	& 24.00 (4.55)	& 17.15 (1.19)	& 28.09 (4.91) & 8.30 & 1.10 \\
SN2007bi    	& 27.00 (14.22) & 9.46 (1.32)	& 25.19 (14.29) & 2.40 & 0.90 \\
SN2010gx   	& 21.18 (0.24)	& 12.97 (0.50) & 31.12 (0.50) & 3.47   &   5.17 \\
SN2010kd   	& 59.00 (3.15)  & 12.35 (1.13) & 60.76 (5.52) & 0.17   & 8.50 \\
\enddata 
\tablecomments{The numbers in parentheses indicate the corresponding 1-$\sigma$ uncertainties of the fitting
parameters. Parameters without cited errors indicate derived physical parameters.$^{a}$ $\chi^{2}$ per
degree of freedom ($dof = N_{d}?N_{p}?1$, where $N_{d}$ is the number of data points and $N_{p}$ is the number
of fitting parameters).$^{b}$ Using Equation 3 and assuming $v =$~10,000~km~s$^{-1}$, $\kappa =$~0.33~cm$^{2}$~g$^{-1}$.}
\end{deluxetable}

\setcounter{table}{3}
\begin{deluxetable}{lcccccccccc}
\tabletypesize{\tiny\tiny\tiny}
\tablewidth{0pt}
\tablecaption{Best-fit parameters for the MAG model.}
\tablehead{
\colhead {SN} &
\colhead {$t_{rise}$~(d)} &
\colhead {$E_{p}$~($10^{51}$~erg)} &
\colhead {$t_{p}$~(d)} &
\colhead {$t_{d}$~(d)} &
\colhead {$R_{0}$~(10$^{13}$~cm)} &
\colhead {$\chi^{2}/dof$} &
\colhead {$M_{ej}^{b}$~($M_{\odot}$)} &
\colhead {$P_{i}$~(ms)} &
\colhead {$B$~($10^{14}$~G)} &\\}
\startdata
SN2008iy		&292.00 (2.28) &1.39 (0.05)& 101.90 (10.48)& 342.20 (6.74)& 31.20 (15.81)& 0.40 &164.13 &3.79& 0.62\\
PTF09uj      	&5.00 (0.08) &0.02 (0.01)& 5.00 (0.16) &4.36 (0.12) &0.01 (0.37) &0.05& 0.03& 29.17& 28.42\\
\hline
SN2006gy    	&49.06 (0.21)& 4.10 (0.10) &12.67 (0.06)& 65.49 (0.02) &2.00 (0.80)& 12.78& 6.01& 2.21& 1.35\\
SN2006tf     	&65.00 (1.09)& 1.09 (0.01) &58.91 (1.40) &58.92 (0.60)& 2.49 (1.68)& 2.83& 4.87 &4.28 &1.22\\
SN2008am  	&41.03 (0.03) &4.07 (0.02) &134.40 (1.01) &26.88 (0.05)& 0.11 (0.22) &1.58& 1.01& 2.22& 0.42\\
SN2008es   	&20.00 (0.002) &2.43 (0.01) &47.31 (0.36) &14.21 (0.09)& 0.04 (0.22)& 2.87& 0.28& 2.87& 0.91\\
CSS100217	&112.00 (2.50) &17.18 (6.44)& 216.10 (23.70) &82.24 (7.59) &85.26 (8.98) &0.78 &9.45 &1.08& 0.16\\
\hline	
SN2005ap   	&16.00 (0.69)& 2.12 (0.07)& 28.87 (4.29) &12.12 (1.20)& 2.46 (0.91) &0.58& 0.21& 3.10& 1.25\\
SCP06F6     	&28.70 (0.59)& 3.21 (0.18)& 9.59 (1.23) &38.63 (1.34) &32.46 (14.77) &3.09& 2.09 &2.50& 1.76\\
SN2007bi    	&61.00 (0.87) &2.79 (0.04)& 19.46 (0.60)& 72.51 (0.70)& 2.36 (0.90)& 2.25 &7.37 &2.67& 1.32\\
SN2010gx   	&14.00 (0.20) &1.49 (0.03) &1.01 (0.01) &35.22 (0.18) &9.97 (0.28) &0.30& 1.74 &3.66 &7.98\\
SN2010kd   	&47.99 (1.63) &2.66 (0.10) &8.06 (1.80) &70.20 (3.50) &2.55 (1.10) &0.13 &6.91 &2.74& 2.10\\
\enddata 
\tablecomments{See comments for Table~3.}
\end{deluxetable}

\setcounter{table}{4}
\begin{deluxetable}{lccccccc}
\tabletypesize{\tiny\tiny\tiny}
\tablewidth{0pt}
\tablecaption{Best-fit parameters for the TH model.}
\tablehead{
\colhead {SN} &
\colhead {$t_{rise}$~(d)} &
\colhead {$E_{sh}$~($10^{51}$~erg)} &
\colhead {$t_{sh}$~(d)} &
\colhead {$t_{d}$~(d)} &
\colhead {$\chi^{2}/dof$} &
\colhead {$M_{CSM,th}^{b}$~($M_{\odot}$)} &\\}
\startdata
SN2008iy		&31.00 (1.63) &0.74 (0.02)& 30.48 (26.55) &461.70 (8.89) &1.92& 298.77\\
PTF09uj      	&5.00 (0.11) &0.01 (0.01)& 4.68 (0.21) &8.37 (0.24) &0.25& 0.09\\
\hline
SN2006gy    	&60.00 (0.08) &1.34 (0.01)& 60.55 (0.51) &51.36 (0.25)& 13.72& 3.70\\
SN2006tf     	&14.00 (5.23) &0.74 (0.07) &13.80 (11.57)& 95.13 (0.52)& 3.80& 12.68\\
SN2008am  	&37.00 (1.63) &2.37 (0.02) &36.80 (2.14) &103.70 (1.29) &1.52 &15.07\\
SN2008es   	&34.00 (11.13)& 1.57 (0.10) &34.16 (15.29) &42.29 (1.86)& 1.28 &2.51\\
CSS100217	&118.20 (0.65) &11.92 (0.08) &83.58 (0.94)& 248.20 (2.50)& 1.55 &86.34\\
\hline	
SN2005ap   	&12.12 (4.68) &1.32 (0.09) &21.5 (5.16) &29.10 (1.35)& 0.54& 1.17\\
SCP06F6     	&43.01 (0.45) &1.15 (0.02) &45.50 (0.75)& 27.94 (1.25)& 1.65& 2.90\\
SN2007bi    	&2.00 (28.20) &0.99 (0.06)& 48.80 (7.30)& 70.15 (0.59) &2.60 &6.90\\
SN2010gx   	&47.00 (0.34) &0.45 (0.01) &46.93 (0.39) &20.32 (0.18) &1.36& 0.58\\
SN2010kd   	&69.00 (5.55) &0.73 (0.02) &69.27 (6.89) &52.39 (0.74) &0.27 &3.85\\
\enddata 
\tablecomments{See comment for Table~3.}
\end{deluxetable}


\setcounter{table}{5}
\begin{deluxetable}{lcccccc}
\tabletypesize{\tiny\tiny\tiny}
\tablewidth{0pt}
\tablecaption{Summary of the fitting parameters for the CSM+RD SN LC model to the hydrogen-rich events.}
\tablehead{
\colhead {Parameter} &
\colhead {SN~2008iy} &
\colhead {PTF~09uj ($s =$~0)} &
\colhead {PTF~09uj ($s =$~2)} &
\colhead {SN~2006gy ($s =$~0)} &
\colhead {SN~2006gy ($s =$~2)} &\\}
\startdata
 $t_{rise}$~(d)							         &241.00 (4.64)& 6.00 (0.19)& 5.00 (0.10)&67.12 (0.09)& 29.00 (0.04)\\
$v_{SN}$~(10$^{3}$~km~s$^{-1}$)  			&5.99 (0.81) &21.46 (0.27) &13.33 (0.17)&35.24 (0.11)& 31.57 (0.47)\\
 $M_{ej}$~($M_{\odot}$)						&36.31 (8.65)& 1.52 (0.08)& 37.39 (2.33)&10.70 (0.08) &7.87 (31.17)\\
$R_{p}$~(10$^{13}$~cm)						&42.89 (5.65) &8.67 (0.55) &32.59 (1.67)&53.12 (3.83) &1.57 (0.10)\\
$M_{CSM}$~($M_{\odot}$)					&38.58 (1.66) &0.03 (0.01) &0.16 (0.01)&5.18 (0.15) &3.64 (0.07)\\
$\rho_{CSM,1}$~(10$^{-12}$~g~cm$^{-3}$)   	          &  0.29(0.14) &99.37(23.93) &1.57(0.29)&0.04 (0.01)& 573.00 (282.00)\\
$M_{Ni}$~($M_{\odot}$)						 & 2.44 (0.26)& 0.28 (0.04) &0.32 (0.10)&3.25 (0.04) &3.80 (0.50)\\
$\chi^{2}/dof$     						           &  0.35& 0.05& 0.05&4.87 &49.30\\
$E_{SN}$~($10^{51}$~erg) 					  &  0.17& 0.09& 0.85&1.70& 1.00\\
 $M_{CSM,th}$~($M_{\odot}$)					 & 38.37& 0.03& 0.15&4.97& 3.45\\
$R_{ph}$~(10$^{14}$~cm)				          & 40.06 &0.93& 4.74&40.51& 37.55\\
 $R_{CSM}$~(10$^{14}$~cm)  			                   & 40.13 &0.93& 4.76&41.07 &39.60\\
$\tau_{CSM}$						                  &   337 &272 &53&44& 2960\\
$t_{RS,*}$~(d)						                  &  458.61& 6.34& 505.86&104.68& 51.23\\
$t_{FS,BO}$~(d)						         & 438.47 &0.07& 4.84&66.98 &86.91\\
\enddata 
\tablecomments{For the CSM+RD best-fit models we have adopted $\delta =$~2 and $n =$~12.}
\end{deluxetable}

\setcounter{table}{5}
\begin{deluxetable}{lccccc}
\tabletypesize{\tiny\tiny\tiny}
\tablewidth{0pt}
\tablecaption{Continued.}
\tablehead{
\colhead {Parameter} &
\colhead {SN~2006tf} &
\colhead {SN~2008am} &
\colhead {SM~2008es} &
\colhead {CSS100217} &\\}
\startdata
$t_{rise}$~(d)							&45.00(1.19) &28.99 (1.30) &21.00 (0.21) & 71.34 (0.86)\\
$v_{SN}$~(10$^{3}$~km~s$^{-1}$)  		&27.08 (0.82) &45.29 (0.98)& 61.85 (0.62)& 35.46 (0.10)\\
$M_{ej}$~($M_{\odot}$)					&33.58 (1.65) &47.77 (7.58) &11.90 (0.33)& 100.2 (0.4)\\
$R_{p}$~(10$^{13}$~cm)  				&36.62 (2.92) &36.56 (3.57) &45.28 (3.40) &99.94 (0.85)\\
$M_{CSM}$~($M_{\odot}$)				&4.72 (0.33) &4.90 (0.30) &2.69 (0.12)  &78.27 (0.65)\\
$\rho_{CSM,1}$~(10$^{-12}$~g~cm$^{-3}$)     &3.15 (0.80) &3.70 (1.55) &0.86 (0.23) & 0.15(0.01)\\
$M_{Ni}$~($M_{\odot}$)       				&0.00 (0.04) &1.57 (1.30) &0.04 (0.06)  &0.50 (0.43)\\
$\chi^{2}/dof$    						&  3.80& 1.52 &1.27&  1.15\\
$E_{SN}$~($10^{51}$~erg) 				& 3.15 &12.54& 5.82& 83.58\\
$M_{CSM,th}$~($M_{\odot}$)				&4.71& 4.89& 2.67&  77.42\\
$R_{ph}$~(10$^{14}$~cm)				&9.14 &8.74 &12.21&  63.44\\
$R_{CSM}$~(10$^{14}$~cm)  				&9.15 &8.74 &12.24 & 63.67\\
$\tau_{CSM}$							&571 &620& 218 & 266\\
$t_{RS,*}$~(d)							&44.83 &28.37& 22.53  &73.13\\
$t_{FS,BO}$~(d)						&10.56 &5.61 &6.94&  56.36\\
\enddata 
\end{deluxetable}

\setcounter{table}{6}
\begin{deluxetable}{lcccc}
\tabletypesize{\tiny\tiny\tiny}
\tablewidth{0pt}
\tablecaption{Summary of the fitting parameters for the CSM+RD SN LC model to the hydrogen-poor events.}
\tablehead{
\colhead {Parameter} &
\colhead {SN~2005ap} &
\colhead {SCP~06F6} &
\colhead {SN~2007bi} &\\}
\startdata
$t_{rise}$~(d)							&15.00 (0.94) &39.39 (0.94)& 59.00 (1.25)\\
$v_{SN}$~(10$^{3}$~km~s$^{-1}$)  		&72.24 (1.60) &43.46 (0.64) &29.96 (0.90)\\
$M_{ej}$~($M_{\odot}$)					&7.36 (0.72) &7.55 (0.39) &44.27 (2.61)\\
$R_{p}$~(10$^{13}$~cm)					& 27.48 (3.84) &34.98 (7.92) &31.28 (4.98)\\
$M_{CSM}$~($M_{\odot}$)				&1.19 (0.07) &4.08 (0.22) &4.28 (0.26)\\
$\rho_{CSM,1}$~(10$^{-12}$~g~cm$^{-3}$)	&1.07(0.68) &0.05(0.03) &1.34(1.21)\\
$M_{Ni}$~($M_{\odot}$)					&3.56 (1.48)& 0.00 (0.28) &0.57 (0.22)\\
$\chi^{2}/dof$							& 0.53 &0.61 &2.60\\
$E_{SN}$~($10^{51}$~erg) 				&4.91& 5.11& 5.08\\
$M_{CSM,th}$~($M_{\odot}$)				&1.18& 3.84& 4.26\\
$R_{ph}$~(10$^{14}$~cm) 				&8.18& 33.35& 11.55\\
$R_{CSM}$~(10$^{14}$~cm)				&8.20 &34.03& 11.57\\
$\tau_{CSM}$							&117& 30.5& 227\\
$t_{RS,*}$~(d)							&14.53 &49.10& 59.00\\
$t_{FS,BO}$~(d)   						&4.12 &34.03 &14.97\\
\enddata 
\tablecomments{Same as Table~6.}
\end{deluxetable}

\setcounter{table}{6}
\begin{deluxetable}{lccccccc}
\tabletypesize{\tiny\tiny\tiny}
\tablewidth{0pt}
\tablecaption{Continued.}
\tablehead{
\colhead {Parameter} &
\colhead {SN~2010gx ($s =$~0)} &
\colhead {SN~2010gx ($s =$~2)} &
\colhead {SN~2010kd} &\\}
\startdata
$t_{rise}$~(d)			    				 &43.00 (0.37)     &31.00 (0.16)   & 58.00 (1.10)	\\    
$v_{SN}$~(10$^{3}$~km~s$^{-1}$)  	          &37.24 (0.68)     &30.37 (0.11)   & 30.01 (0.57)	\\	  
$M_{ej}$~($M_{\odot}$)					     &17.59 (0.78)     &9.70 (0.26)    & 34.10 (1.91)	\\    
$R_{p}$~(10$^{13}$~cm)		   			  &93.50 (5.07)      &2.00 (1.51)    & 43.25 (5.31)	\\    
$M_{CSM}$~($M_{\odot}$)		  		   &1.39 (0.05)      &1.64 (0.26)    & 3.28 (0.15)	\\    
$\rho_{CSM,1}$~(10$^{-12}$~g~cm$^{-3}$)      &0.03 (0.01)  &1664.30 (975.90)& 1.09 (0.66)\\    
$M_{Ni}$~($M_{\odot}$)		  			   &0.00 (0.01)      &0.00 (0.01)    & 0.00 (0.15)	\\    
$\chi^{2}/dof$~$^{b}$                 				 &1.88	       &0.98         & 0.27 	\\    
$E_{SN}$~($10^{51}$~erg)  	     			&3.12	       &1.14	       & 3.93		\\    
$M_{CSM,th}$~($M_{\odot}$)	   			  &1.37	       &1.61	       & 3.26		\\    
$R_{ph}$~(10$^{14}$~cm)		    		 &12.00	       &15.50	       & 11.47  	\\    
$R_{CSM}$~(10$^{14}$~cm)  	 			   &12.00	       &15.80	       & 11.49  	\\    
$\tau_{CSM}$							     &2	       &6573	       & 157		\\    
$t_{RS,*}$~(d)							     &42.88	       &57.93	       & 57.90  	\\    
$t_{FS,BO}$~(d)			   			  &2.64	       &33.37	       & 12.07  	\\    
\enddata 
\end{deluxetable}

\setcounter{table}{7}
\begin{deluxetable}{lcccccc}
\tabletypesize{\tiny}
\tablewidth{0pt}
\tablecaption{Summary of the characteristics of the sample of transients studied in this work.}
\tablehead{
\colhead {Event} &
\colhead {Type} &
\colhead {Spectral properties} &
\colhead {Best-fit LC model$^{a}$} &\\}
\startdata
SN~2008iy    &IIn     &  Pure H emission, weak P Cygni  &   CSM+RD/TH			     \\
PTF~09uj     &IIn     &  Pure H emission  weak P Cygni  &   CSM+RD/TH			     \\
\hline
SN~2006gy    &IIn    &  P Cygni H			&   CSM+RD/TH			     \\
SN~2006tf    &IIn     &  Pure H emission, weak P Cygni  &   CSM+RD/TH			     \\
SN~2008am    &IIn     &  Pure H emission		&   CSM+RD/TH			     \\
SN~2008es    &II-L?   &  P Cygni H			&   MAG, CSM+RD 		     \\
CSS100217    &IIn     &  Pure H emission, weak P Cygni  &   CSM+RD/TH			     \\
\hline
SN~2005ap    &II-L    &  Broad P Cygni C, N, O  	&   CSM+RD/TH			     \\
SCP06F6      &?       &  Peculiar			&   CSM+RD/TH			     \\
SN~2007bi    &Ic      &  Ni/Co/Fe blends, no H  	&   RD, MAG, CSM+RD/TH  	     \\
SN~2010gx    &Ib/c      &  P Cygni CaII, FeII, SiII	&   CSM+RD/TH			     \\
SN~2010kd    &Ic      &  Peculiar, no H, He		&   MAG, CSM+RD/TH		     \\
\enddata 
\tablecomments{$^{a}$ RD = $^{56}$Ni and $^{56}$Co radioactive decay diffusion model, MAG = Magnetar spin-down model, TH = ``Top-hat" model and 
CSM+RD = Hybrid SN ejecta-CSM interaction plus $^{56}$Ni and $^{56}$Co decay model, CSM = SN ejecta-CSM interaction model (zero input
from radioactive decays).}
\end{deluxetable}


\setcounter{table}{0}
\begin{deluxetable}{llllllccccc}
\tabletypesize{\tiny}
\renewcommand\thetable{A1}
\tablewidth{0pt}
\tablecaption{Model parameters and correlation coefficients for the RD model. 1-$\sigma$ errors are given in parentheses.}
\tablehead{
\colhead {} &
\colhead {$t_{exp}$} &
\colhead {$M_{Ni}$} &
\colhead {$t_{d}$} &
\colhead {$A_{\gamma}$}\\}
\startdata
initial   &   0.0     &   10.0   &  60.0 &      300.0 \\
recovered & 0.0       &10.0       & 60.0        & 300.0  \\
          & (0.17)    &(0.04)     & (0.32)      & (3.11) \\
\hline
$t_{exp}$ & 1.00     & -0.67      & -0.86       & 0.17  \\ 
$M_{Ni}$  &          & 1.00       & 0.91        & -0.74 \\ 
$t_d$     &          &            & 1.00        & -0.54 \\ 
\enddata 
\end{deluxetable}

\setcounter{table}{0}
\begin{deluxetable}{llllllccccc}
\tabletypesize{\tiny}
\renewcommand\thetable{A2}
\tablewidth{0pt}
\tablecaption{Same as Table A1 but for the MAG model.}
\tablehead{
\colhead {} &
\colhead {$t_{exp}$} &
\colhead {$R_{0}$} &
\colhead {$E_{p}$} &
\colhead {$t_{d}$} &
\colhead {$t_{p}$} &
\colhead {$v$} &
\\}
\startdata
initial &  0.0 & 1.0 & 2.0 & 50.0 & 30.0 & 40.0 \\
recovered & 0.0  & 0.47  & 2.00  & 49.99  & 30.07  & 42.88 \\
          &(0.03)& (0.16)& (0.01)& (0.19) & (0.30) & (10.63)\\
\hline
$t_{exp}$ & 1.00& -0.47 &0.12& -0.03& -0.24& -0.44\\
$R_0$ &  & 1.00& -0.39& -0.17& 0.31& 0.33\\
$E_p$ & & & 1.00& 0.91& -0.91& -0.20 \\
$t_d$ & & & & 1.00& -0.91& -0.19 \\
$t_p$ & & & & & 1.00& 0.31 \\
\enddata 
\end{deluxetable}

\setcounter{table}{0}
\begin{deluxetable}{llllllccccc}
\tabletypesize{\tiny}
\renewcommand\thetable{A3}
\tablewidth{0pt}
\tablecaption{Same as Table A1 but for the TH model.}
\tablehead{
\colhead {} &
\colhead {$t_{exp}$} &
\colhead {$E_{sh}$} &
\colhead {$t_{sh}$} &
\colhead {$t_{0}$} 
\\}
\startdata
initial &  0.0 & 1.0 & 30.0 & 40.0 \\
recovered & 0.00  & 1.00   & 30.00  & 40.00  \\
          & (0.04)& (0.001)& (0.07) & (0.08) \\
\hline
$t_{exp}$ & 1.00 & -0.19 & -0.79 & 0.17 \\ 
$E_{sh}$ & & 1.00 & -0.09 & 0.60 \\ 
$t_{sh}$ & &  & 1.00 & -0.50 \\ 
\enddata 
\end{deluxetable}
 
\setcounter{table}{0}
\begin{deluxetable}{llllllccccccc}
\tabletypesize{\tiny}
\renewcommand\thetable{A4}
\tablewidth{0pt}
\tablecaption{Model parameters and correlation coefficients for the CSM+RD model.}
\tablehead{
\colhead {} &
\colhead {$t_{exp}$} &
\colhead {$R_{0}$} &
\colhead {$M_{ej}$} &
\colhead {$M_{CSM}$} &
\colhead {$\dot{M}$} &
\colhead {$M_{Ni}$} &
\colhead {$v_{SN}$} 
\\}
\startdata
initial &  0.0 & 33.0 & 35.0 & 4.0 & 0.01 & 5.0 & 40.0 \\
recovered $s =$~0 & 0.20 &42.80 &32.72 &4.29& 0.40& 4.67& 41.15\\
                  & (0.02)& (0.44)& (0.09)& (0.02) &(0.01) &(0.03) &(0.05)\\
\hline
$t_{exp}$ & 1.00 &0.40& -0.35 &0.06& -0.21& 0.06& 0.51\\
$R_0$ &  & 1.00 &-0.90& 0.73& -0.89& -0.58& 0.97\\
$M_{ej}$ & & & 1.00& -0.69& 0.87& 0.54 &-0.89 \\
$M_{CSM}$ & & & & 1.00 &-0.80 &-0.88& 0.68\\
$\dot{M}$ & & & & & 1.00 &0.63& -0.86\\
$M_{Ni}$ & & & & & & 1.00& -0.54\\
\hline
recovered $s =$~2  & 3.85 &17.23 &1.11 &4.68 &10.33 &6.90& 40.00\\
                   & (0.07)& (1.69) &(0.07) &(0.04) &(2.08) &(0.06) &(0.13)\\
\hline
$t_{exp}$ & 1.00& -0.62 &0.44& -0.84 &0.63& 0.01& -0.40\\
$R_0$ &  & 1.00 &-0.42& 0.80& -0.98 &0.10& 0.21 \\
$M_{ej}$ & & & 1.00& -0.57 &0.45& -0.63& -0.86 \\
$M_{CSM}$ & & & & 1.00 &-0.79 &0.06 &0.54 \\
$\dot{M}$ & & & & & 1.00 &-0.15& -0.23\\
$M_{Ni}$ & & & & & & 1.00& 0.25\\
\enddata 
\end{deluxetable}


\end{document}